\documentclass[letterpaper,11pt,leqno]{article}
\usepackage{paper}
\usepackage{math}
\usepackage{dsfont}
\usepackage{nicefrac}

\usepackage{tikz}
\usepackage{pgfplots}
\pgfplotsset{compat=1.7}
\usepgfplotslibrary{groupplots}
\usepackage{pgfplotstable}

\hypersetup{pdftitle={Financing Staged Entry}}

\newtheorem{prop}{Proposition}
\newtheorem{ass}{Assumption}

\newcommand{\bib}{bibliography.bib}

\newcommand{\Prob}{\mathbb{P}\mathrm{rob}}

\begin{document}

\title{Staged Entry}

\author{Francesco Del Prato, Paolo Zacchia
%
\thanks{Francesco Del Prato: \emph{Department of Economics and Business Economics, Aarhus University}; e-mail address: \href{mailto:francesco.delprato@econ.au.dk}{francesco.delprato@econ.au.dk}. Paolo Zacchia: \emph{Department of Economics, Ca' Foscari University of Venice; CERGE-EI; and CEPR}; e-mail address: \href{mailto:paolo.zacchia@unive.it}{paolo.zacchia@unive.it}. We thank Daisuke Adachi, Paolo Martellini and Jose Vasquez for helpful comments. A restricted and incomplete version of the model presented in this paper has previously circulated as part of the (currently dormant) paper titled “The Heterogeneous Consequences of Reduced Labor Costs on Firm Productivity.” A previous version of this paper has also circulated under the title “Gatekeeping, Selection and Welfare.”}}

\date{July 2026}


\begin{titlepage}
\maketitle

We develop a model of staged entry: to operate, monopolistically competitive firms must pay two sequential entry costs, each time acquiring a more informative signal of future performance. This model yields two implications for fiscal policy. First, the equilibrium outcome is constrained-efficient if preferences are CES and entry costs are exogenous. Second, when entry costs depend on how many firms pass any entry stage (due to positive knowledge spillovers or negative congestion effects) the resulting externalities can be offset via Pigouvian taxes or subsidies that are timed around the relevant entry decisions. A calibration exercise based on U.S. firm-entry data shows that these policies would raise welfare through a wider pool of entrants and sharper selection of productive firms.

\vspace{1em}
{\footnotesize{\color{gray}\textsc{Keywords}}: staged entry, monopolistic competition, taxation, welfare.}\\
{\footnotesize{\color{gray}\textsc{JEL Classification}}: D61, D62, L11, L26.}

\end{titlepage}

\section{Introduction}\label{sec:introduction}

Firm entry often proceeds in stages. Regulators license firms before market access, banks must satisfy supervisory stress tests before expanding balance sheets, and drug developers must pass increasingly demanding trials before commercialization. At each stage, decision-makers act on imperfect information: they acquire noisy indicators of future performance, which is only revealed after resources are committed \citep{jovanovic1982selection}. Proceeding through the stages entails costs that may be influenced by other firms. If, for example, \emph{knowledge spillovers} occur, entrants may learn from other developing or existing firms, lowering those costs. By contrast, \emph{congestion forces} operate if those costs increase as the number of entrants rises, due to e.g. a restricted access to financing. Multiple governments have introduced fiscal instruments affecting incentives for business creation and entrepreneurship.\footnote{These instruments typically take the form of subsidies, tax credits, subsidized loans, loan guarantees, or public venture funding targeted at startups and innovative firms. Examples include R\&D tax credits in France and Canada, Italy’s program for innovative startups, and government-backed credit guarantee schemes widely used across OECD countries to ease financing constraints for young firms.} Under what conditions are these policies welfare-enhancing?

This paper addresses this question with two related contributions. First, we develop a model of \emph{staged entry} in which business ideas develop by accumulating \emph{signals} about future firm productivity and performance. The model extends the monopolistic-competition framework by \cite{melitzImpactTradeIntraIndustry2003} in a closed economy: at every stage, potential firms must pay a stage-specific, fixed, sunk \emph{entry cost} in order to observe a new signal, or final productivity.\footnote{For ease of exposition, we only entertain a model with two entry stages and a final production stage: they are sufficient to deliver the key intuitions. Adding more entry stages is conceptually easy but laborious.} The model characterizes a (self-)selection mechanism whereby only those entrants that observe sufficiently promising signals advance to the successive stage. We initially focus on a version of the model where all entry costs are exogenous: we show that a steady-state equilibrium exists, is unique, and is also constrained-efficient, yielding the allocation that a social planner with the same information as decentralized firms would choose. This result, which extends \cite{dhingraMonopolisticCompetitionOptimum2019}, implies that policymakers should abstain from intervening in firm entry, and serves as a benchmark for evaluating social welfare under variations and extensions.\footnote{This lets us isolate a different source of inefficiency from the non-CES demand effects and selection failures emphasized by \citet{zhelobodko2012monopolistic}, \citet{behrens2020quantifying}, and \citet{bagwell2023monopolistic}: even under CES, entry is inefficient when the cost of reaching a stage depends on how many firms reach earlier stages.}

The second contribution concerns specific departures from the benchmark model that motivate policy intervention. We endogenize entry costs, making them functions of the equilibrium mass of firms surviving each stage of the entry process. This extension can capture real-world phenomena such as the spillover and congestion forces mentioned initially. Predictably, externalities dismantle the benchmark constrained-efficiency result, as private and social costs of entry diverge. We examine two illustrative scenarios: in one, spillover and congestion forces depend on all initial entrants; in the other, only on those paying the corresponding costs. We show that in general, Pigouvian subsidies and taxes re-align private with social incentives. However, they must be timed when the relevant firms choose whether to advance to the next stage, not when the stage-specific entry costs are paid: intuitively, a planner shall anticipate all downstream, later-stage effects that ``advancing'' firms generate. Although the optimal instruments are \emph{budget-neutral} only in knife-edge cases,\footnote{We call a combination of Pigouvian fiscal instruments \emph{budget-neutral} if it features both taxes and subsidies in equivalent amounts; thus, it requires no potentially distortionary interventions elsewhere in the economy. In our model, post-entry taxes and subsidies (assigned at the production stage) are distortionary: as in \citet{dhingraMonopolisticCompetitionOptimum2019}, this follows from consumer preferences with Constant Elasticity of Substitution (CES).} they set incentives for attempting entry in the precise amount that is necessary to use the available resources efficiently and attain a social optimum.

We perform a quantitative exercise to illustrate the potential welfare gains of policy instruments that align private and social incentives. We calibrate our model using U.S. Census data on firm entry, interpreted as the outcome of a decentralized equilibrium. With isoelastic entry costs and under multiple hypotheses on the magnitude of spillover and congestion forces, we identify the welfare gains induced in social optima by optimal Pigouvian instruments. In our central scenarios, these gains amount to slightly less than 1 percent, and increase along the strength of entry spillovers. These gains hinge partly on a variety effect (more active firms) and to a larger extent on higher average productivity, which is driven by tougher competition and tighter selection among more entrants. Our theory accommodates multiple equilibria, allowing for possibly stronger welfare gains unaccounted for by our quantitative exercise. We develop an example where the economy sits on an inferior equilibrium among many, providing a rationale for ``big push'' policies spurring entrepreneurship.

Our policy results speak to the welfare economics of firm entry and entrepreneurship policy. Since at least \citet{spence1976product} and \citet{mankiw1986free}, a central question has been whether private incentives for entry match social incentives. More recently, and closer to our policy question, \citet{melcangi2023subsidizing} show that lump-sum transfers to startups can amplify inefficient entry and misallocation when informational frictions make entry and hiring decisions error-prone. By contrast, we develop a framework where inefficiencies stem from firm-to-firm externalities on entry costs. Imperfect information still plays a role in shaping our policy results: because an attempting entrant's survival at later stages is uncertain (and so is its ability to pay taxes, or enjoy subsidies), fiscal policy should be timed at those stages when the relevant externalities are shaped. Throughout the paper, we emphasize how the ``congestion'' externalities that our model accommodates can be interpreted in terms of imperfect financial markets. This relates our contribution to extant work on credit frictions and firm selection, including \citet{townsend1979optimal}, \citet{manovaCreditConstraintsHeterogeneous2013}, \citet{chaneyLiquidityConstrainedExporters2016}, and \citet{ungerCreditFrictionsSelection2021}.

This paper proceeds as follows. Section \ref{sec:setup} describes the setup of the model.
Section \ref{sec:benchmark} illustrates key results about the benchmark model with exogenous entry costs. Section \ref{sec:externalities} assesses the fiscal policy implications of endogenous entry costs (externalities). Section \ref{sec:calibration} calibrates the benchmark model on U.S.\ firm-entry data and quantifies the mechanisms behind the welfare gains from Pigouvian instruments. Lastly, Section \ref{sec:conclusion} concludes. An appendix collects mathematical proofs and details about the calibration.

\section{Setup}\label{sec:setup}

We study a closed economy populated by a representative consumer whose preferences for individual goods (varieties) feature Constant Elasticity of Substitution (CES). The utility function is expressed as $U^{\frac{\sigma-1}{\sigma}}=\int_{\omega\in\Omega}q\left(\omega\right)^{\frac{\sigma-1}{\sigma}}d\omega$, where $\Omega$ is the set of product varieties available in equilibrium, $q\left(\omega\right)$ denotes the quantity of product $\omega\in\Omega$ consumed, and $\sigma>1$ is the elasticity of substitution. The firms that supply the varieties in $\Omega$ feature heterogeneous \emph{productivity} $\varphi\left(\omega\right)>0$. Their cost function is linear in the labor input $l$ and, due to operational fixed costs $f>0$, exhibits increasing returns to scale. The labor demand function is thus linear in quantity and (dropping dependence on $\omega$ from the notation) is expressed as $l\left(q\right)=f+q/\varphi$. In this economy, labor is inelastically supplied by a mass of workers $L$. Each unit of labor receives a wage $w$, normalized to unity ($w=1$).

This economy inherits the standard properties of the monopolistic competition model by \cite{dixitMonopolisticCompetitionOptimum1977} as extended by \cite{melitzImpactTradeIntraIndustry2003}. In particular, for each firm optimal quantity is a power function of productivity with exponent $\sigma$, whereas both revenue and profit scale with exponent $\sigma-1$. Hence, for any two firms with productivity $\varphi_{1}$ and $\varphi_{2}$, the ratio of their equilibrium revenues $r(\varphi)$ is $r(\varphi_{1})/r(\varphi_{2})=\left(\varphi_{1}/\varphi_{2}\right)^{\sigma-1}$. As in \cite{melitzImpactTradeIntraIndustry2003}, the probability distribution of productivity in this model is endogenous and determined by competitive selection of firms in equilibrium. However, the details of the protocol differ in our model: firms engage in a \emph{three-stages entry} process, which proceeds as follows.

\begin{enumerate}[label=\arabic*., itemsep=0.5\baselineskip, topsep=0.5\baselineskip]
	\item A given mass of potential firms, $M_{e}$, attempts entry. Each entrant must incur a sunk, one-time \emph{experimentation cost} $f_{n}>0$, measured in labor units. Paying it reveals a \emph{signal} $\theta$ about each firm's true productivity $\varphi$.
	\item Having paid $f_{n}$ and observed $\theta$, potential entrants decide whether to incur another sunk one-time \emph{setup cost} $f_{b}>0$, again measured in labor units. This allows potential entrants to observe their \emph{true} productivity $\varphi$.
	\item Firms that pay both sunk costs can engage in production, setting prices and quantities. They can exit and supply zero output if, due to operational fixed costs $f$, optimal profits conditional on producing are negative. Firms that stay operate in the economy until some event that occurs with an exogenous probability $\delta\in\left(0,1\right)$ forces them to exit.
\end{enumerate}

In steady state, this process yields an equilibrium mass of firms $M$ that supplies the varieties in $\Omega$ available to consumers. We further depart from \cite{melitzImpactTradeIntraIndustry2003} by allowing both sunk entry costs to be functions of (endogenous) features of the model. Whereas more general formulations are possible, this paper focuses on two cases: \begin{enumerate*}[label=\roman*., itemjoin={{; }}, itemjoin*={{; and }}] \item both $f_{n}$ and $f_{b}$ are fixed and exogenous \item they are continuously differentiable functions of endogenous variables such as $M_{e}$, expressed e.g.\ as $f_{n}=f_{n}\left(M_{e}\right)$ and $f_{b}=f_{b}\left(M_{e}\right)$\end{enumerate*}. The latter case can capture relevant economic mechanisms and externalities. One can, for example, express both \emph{knowledge spillovers} at the experimentation stage and \emph{screening congestion} under the restrictions that $f_{n}^{\prime}\left(M_{e}\right)<0$ and $f_{b}^{\prime}\left(M_{e}\right)>0$ for any value (or monotone function) of $M_{e}$.

The signal $\theta$ and productivity $\varphi$ are, for each potential entrant, drawn from a joint probability distribution governed by the following assumptions.

\begin{ass}\label{ass:A1}\normalfont\emph{Regular distributions.} The signal $\theta$ has a continuously differentiable and atomless marginal distribution $C\left(\theta\right)$ whose support is a connected interval $\Theta=(\underline{\theta},\overline{\theta})\subseteq\mathbb{R}$, where either endpoint may be infinite. For every $\theta\in\Theta$, there exists a regular, atomless conditional distribution $G\left(\varphi\mid\theta\right)$ of productivity where $\varphi>0$, and the map $\theta\mapsto G\left(\varphi\mid\theta\right)$ is weakly continuous.
\end{ass}
\begin{ass}\label{ass:A2}\normalfont\emph{Finite CES-relevant productivity moments.} Productivity $\varphi$ possesses a finite unconditional uncentered moment of order $\sigma-1$: $\mathbb{E}\left[\varphi^{\sigma-1}\right]<\infty$.
\end{ass}
\begin{ass}\label{ass:A3}\normalfont\emph{Strict signal informativeness.}
If $\theta_1>\theta_2$, then $G(\cdot\mid\theta_1)$ strictly first-order stochastically dominates $G(\cdot\mid\theta_2)$.
\end{ass}
\begin{ass}\label{ass:A4}\normalfont\emph{Signal-responsive productivity tail.} For every $\theta\in\Theta$ and every finite $\varphi>0$, $\Prob([\varphi,\infty)\mid \theta)>0$. Moreover, for every finite $\varphi>0$,
\begin{align*}
\Prob\left(\left[\varphi,\infty\right)\mid \theta\right)\to 0 &\quad\text{as }\theta\downarrow \underline{\theta}, \ \text{and}\\
\Prob\left(\left[\varphi,\infty\right)\mid \theta\right)\to 1 &\quad\text{as }\theta\uparrow \overline{\theta}
\end{align*}
where the two limits are understood as $\theta\to-\infty$ or $\theta\to+\infty$ whenever the corresponding endpoint of $\Theta$ is unbounded.
\end{ass}

Assumption \ref{ass:A1} rules out oddly-defined ``signals'' and is only introduced for tractability's sake. Assumption \ref{ass:A2} is a requirement typical of analogous CES environments: $\varphi^{\sigma-1}$ is finite in expectation, so that firm choices and thus, equilibria are well-defined. Assumption \ref{ass:A3} is key: signals are ordered so that ``better'' ones predict higher productivity throughout the latter's support. It rules out instances where, e.g., signals shift prevalently (or solely) higher moments of $\varphi$. Assumption \ref{ass:A4} establishes that the tail of the productivity distribution is sufficiently thick and that at the endpoints of $\Theta$ one can predict almost deterministically whether $\varphi$ occurs in the tail or not: intuitively, $G\left(\varphi\mid\theta\right)$ becomes increasingly concentrated in the tail, or away from it. These assumptions accommodate multiple familiar cases, such as $\left(\theta,\varphi\right)$ following a bivariate lognormal distribution with positive correlation; and $G\left(\varphi\mid\theta\right)$ being Type-I Pareto-distributed.\footnote{For the Pareto case, the assumptions cover instances where the scale parameter of $G\left(\varphi\mid\theta\right)$ is a suitable positively monotonic function of $\theta$, while the shape parameter is large enough for Assumption \ref{ass:A2} to hold.}

The model is solved recursively. Let $\widetilde{\pi}\left(\theta\right)=\mathbb{E}_{\left.\varphi\right|\theta}\left[\left.\pi(\varphi)\right|\theta\right]$ be the \emph{ex ante} profits that a firm, which observes signal $\theta$, can expect at stage two before paying $f_{b}$ and observing $\varphi$. Here, $\pi\left(\varphi\right)$ denotes the \emph{ex post} stage-three profits as a function of productivity. Adapting the analysis of the Melitz model, $\widetilde{\pi}\left(\theta\right)$ is understood as a function of $\varphi^{\ast}$: the productivity threshold below which firms renounce production, where $\pi\left(\varphi^{\ast}\right)=0$. Specifically:
\begin{equation}\label{eq:pi-tilde}
\widetilde{\pi}\left(\theta;\varphi^{\ast}\right)=f\left\{\int_{\varphi^{\ast}}^{\infty}\left(\frac{\varphi}{\varphi^{\ast}}\right)^{\sigma-1}dG\left(\varphi\mid\theta\right)-\left[1-G\left(\varphi^{\ast}\mid\theta\right)\right]\right\}.
\end{equation}
This expression implicitly embeds a Melitz-style Zero Profit Condition (ZPC), which is specific to signal $\theta$. Under Assumption \ref{ass:A3}, $\widetilde{\pi}\left(\theta;\varphi^{\ast}\right)$ is strictly increasing in $\theta$, since better signals always predict higher productivity (and profits).

This analysis implies the existence of a signal threshold $\theta^{\ast}$ below which firms would not pay $f_{b}$ at stage two. At $\theta^{\ast}$, the present value of expected profits equals the setup cost:
\begin{equation}\label{eq:arbitrage}
\frac{\widetilde{\pi}(\theta^{\ast};\varphi^{\ast})}{\delta}-f_{b}=0.
\end{equation}
For reasons clarified below, we call \eqref{eq:arbitrage} the Arbitrage Condition (AC). The model is closed through an appropriate Free Entry (FE) condition:
\begin{equation}\label{eq:free_entry}
\int_{\theta^{\ast}}^{\overline{\theta}}\frac{\widetilde{\pi}\left(\theta;\varphi^{\ast}\right)}{\delta}dC\left(\theta\right)-\left[1-C\left(\theta^{\ast}\right)\right]f_{b}-f_{n}=0.
\end{equation}
Expression \eqref{eq:free_entry} subsumes all \emph{ex ante}, stage-one incentives: to dissuade further entrants, the expected present value of entry (before signals are observed) must be equal to the sum of the experimentation cost $f_{n}$ and the \emph{expected} setup cost $\left[1-C\left(\theta^{\ast}\right)\right]f_{b}$. In fact, at stage two entrants would only pay $f_{b}$ if their signal exceeds $\theta^{\ast}$.

One can enhance this model's interpretation through additional microfoundations. Suppose, for example, that all setup costs are provided by some risk-neutral intermediaries (``banks'') in exchange for a claim on a share $s\left(\theta\right)$ of firms' future profits.\footnote{Existing versions of the Melitz model that incorporate financial intermediaries or frictions \citep[e.g.,][]{manovaCreditConstraintsHeterogeneous2013,chaneyLiquidityConstrainedExporters2016} typically introduce liquidity constraints that firms face only when they encounter costs for entering foreign markets. This microfoundation introduces them at the entry stage.} In this financial market, banks can verifiably observe signals $\theta$ and discriminate accordingly. If there are many such banks that compete \emph{\`{a} la} Bertrand, $s\left(\theta\right)$ is set so that the expected bank claims equal exactly $f_{b}$ for all $\theta\in\Theta$. It follows that if $\theta<\theta^{\ast}$ entrants are not financed; $s\left(\theta^{\ast}\right)=1$; and $\widetilde{\pi}(\theta)s\left(\theta\right)=\widetilde{\pi}(\theta^{\ast})$ if $\theta>\theta^{\ast}$. Consistently with \eqref{eq:free_entry}, at stage one entering firms expect to return $\left[1-C\left(\theta^{\ast}\right)\right]f_{b}$ to banks \emph{ex post}. This microfoundation motivates the ``arbitrage'' label for \eqref{eq:arbitrage}, as well as stage-two congestion effects (i.e., $f_{b}^{\prime}\left(M_{e}\right)>0$). The latter would occur if, for example, banks' screening and operational costs increase when they face an increasing number of loan applicants (entering firms).

\section{Benchmark}\label{sec:benchmark}

We first analyze the benchmark case where $\left(f_{n},f_{b}\right)$ are fixed and exogenous constants. We begin by characterizing the model's equilibrium as a pair of signal and productivity thresholds $\left(\theta^{\ast},\varphi^{\ast}\right)$, to then analyze its welfare and efficiency implications.

\begin{prop}[Benchmark equilibrium]\label{prop:benchmark} Under Assumptions \ref{ass:A1}--\ref{ass:A4}, there exists a unique interior equilibrium pair $\left(\theta^{\ast},\varphi^{\ast}\right)$, which satisfies the Arbitrage Condition \eqref{eq:arbitrage} and meets the implicit function $\varphi^{\ast}(\theta^{\ast})$ based on the Free Entry condition \eqref{eq:free_entry} at its maximum.
\end{prop}

To provide intuition, Figure \ref{fig:Equilibrium} illustrates the equilibrium as the intersection between the AC and FE curves on the $(\theta,\varphi)$ plane when $\Theta=\mathbb{R}_{++}$. The AC curve increases monotonically because higher signal thresholds sharpen selection and \emph{ex post} competition. The FE curve is instead concave by the interplay of multiple mechanisms. As in the Melitz model, the higher the productivity threshold, the higher the profits required to motivate entry. The signal threshold, however, affects the condition in two ways. A higher value of $\theta^{\ast}$, similarly to $\varphi^{\ast}$, requires higher profits to motivate entry: the first element on the left-hand side of \eqref{eq:free_entry}. At the same time, it also reduces the probability of bearing the setup cost $f_{b}$ after observing the signal (second element), implicitly reducing the productivity threshold that keeps stage-one \emph{ex ante} profits constant. The first mechanism dominates at low values of $\theta$; the second, at higher ones. In equilibrium these two forces are optimally balanced: as the Proposition states and Figure \ref{fig:Equilibrium} shows, the AC curve cuts the FC curve at the latter's maximum. Given $\theta^{\ast}$, the expected profits from entry are maximized in equilibrium.

\begin{figure}[ht!]
	\caption{Benchmark equilibrium and comparative statics}
	\label{fig:Equilibrium}
	\centering
	\begin{tikzpicture}
	    \begin{axis}
	        [axis lines = middle, samples = 100, domain=0:2, x = 3cm, y = 2.6cm,
            xmin = 0, xmax = 2, ymin = 0, ymax = 2, enlargelimits,
            xtick={1,1.2}, xticklabels={$\theta^{\ast}$,$\theta^{\ast\ast}$}, xlabel = $\theta$,
            x tick label style = {font = \small},
            ylabel = $\varphi$, ytick={1.2,1.6734}, yticklabels={$\varphi^{\ast}$,$\varphi^{\ast\ast}$},
            y tick label style = {font = \small}]
	        \addplot [domain = 0:1.75, very thick] {0.95+0.5*x^(2)-0.25*x^(4)};
	        \addplot [domain = 0:1.75, very thick] {1.2*(x)^(0.6)};
	        \addplot [domain = 0:0.20, very thick, dashed] {1.4234-0.5*(x-0.2)^(2)+0.25*(x-0.2)^(4)};
	        \addplot [domain = 0.20:1.75, very thick, dashed] {1.4234+0.5*(x-0.2)^(2)-0.25*(x-0.2)^(4)};
	        \addplot [domain = 0:1.75, very thick, dashed] {1.5*(x)^(0.6)};
	        \draw [densely dotted] (axis cs:1,0) -- (axis cs:1,1.2) -- (axis cs:0,1.2);
	        \draw [densely dotted] (axis cs:1.2,0) -- (axis cs:1.2,1.6734) -- (axis cs:0,1.6734);
	        \node [right] at (axis cs:1.75,0.2) {$\mathrm{FE}$};
	        \node [right] at (axis cs:1.75,1.2) {$\mathrm{FE}^{\prime}$};
	        \node [right] at (axis cs:1.75,1.7) {$\mathrm{AC}$};
	        \node [right] at (axis cs:1.75,2.1) {$\mathrm{AC}^{\prime}$};
	        \end{axis}
	\end{tikzpicture}
	
    \note[Note]{This figure illustrates the benchmark equilibrium $\left(\theta^{\ast},\varphi^{\ast}\right)$ as the intersection between the solid lines representing the AC \eqref{eq:arbitrage} and FE \eqref{eq:free_entry} conditions. The dashed lines represent the new AC and FE curves resulting from increasing operational wages to a higher value.}
\end{figure}

Figure \ref{fig:Equilibrium} also illustrates a relevant comparative statics exercise. Consider an increase in the \emph{operational} labor costs that firms face from stage three onward. Specifically, firms face an increase in the effective wage $w$ paid to production workers $l\left(q\right)$, but the two entry costs $f_{n}$ and $f_{b}$ stay unaffected. This can be consequent to an exogenous tax or wedge that only affects production workers. In this thought experiment, both the AC and FE curves shift or rotate upward, as displayed via the dashed lines of Figure \ref{fig:Equilibrium}. Because of tightened selection at stage three, both thresholds are higher at the new equilibrium $\left(\theta^{\ast\ast},\varphi^{\ast\ast}\right)$.

As in related models, social welfare (i.e., the representative consumer's utility) is equal to the inverse of the price level and can be expressed as $U=\frac{\sigma-1}{\sigma}M^{\nicefrac{1}{\sigma-1}}\widetilde{\varphi}\left(\theta^{\ast},\varphi^{\ast}\right)$, where $M$ is the equilibrium mass of firms while $\widetilde{\varphi}\left(\theta^{\ast},\varphi^{\ast}\right)$ is \emph{aggregate productivity}:
\begin{equation*}
\widetilde{\varphi}\left(\theta^{\ast},\varphi^{\ast}\right)\coloneqq\mathbb{E}\left[\varphi^{\sigma-1}\mid\theta\geq\theta^{\ast},\varphi\geq\varphi^{\ast}\right]^{\smash{\nicefrac{1}{\sigma-1}}}. 
\end{equation*}
The benchmark equilibrium is Pareto-optimal as per the following statement.

\begin{prop}[Benchmark constrained efficiency]\label{prop:optimality}
The decentralized benchmark equilibrium is constrained-efficient: the equilibrium thresholds $\left(\theta^{\ast},\varphi^{\ast}\right)$ solve the welfare maximization problem of a planner who, at every stage, has the same information as potential entrants.
\end{prop}

This result and the associated proof extend \citet{dhingraMonopolisticCompetitionOptimum2019}. In particular, the proof shows that the planner's problem can be solved recursively in stages; at every stage the planner and the firms face the same incentives provided that they also share the same information.\footnote{As in \citet{dhingraMonopolisticCompetitionOptimum2019}, this result collapses if preferences are no longer CES. Using their terminology, CES preferences are a unique case where the \emph{private markup} and the \emph{social markup} coincide for every variety $\omega$; hence, the planner and decentralized firms face the same incentives. To our purposes, CES preferences are a useful benchmark to examine tax instruments under entry externalities, since they isolate our inefficiency sources of interest.} A key implication of this result is that it is suboptimal for a planner to use policy instruments like taxes to mitigate the welfare losses associated with the entry cost spent on firms that ultimately exhibit subpar productivity.

We illustrate the last point with an example about two opposite \emph{limit} cases specializing to more conventional Melitz economies: \begin{enumerate*}[label=\arabic*., itemjoin={{, }}, itemjoin*={{, and }}] \item that where $\theta$ and $\varphi$ are statistically independent \item that where they coincide, that is, $\Theta=\mathbb{R}_{+}$ and $G\left(\varphi\mid\theta\right)$ is a step function whose unique discontinuity occurs at $\varphi=\theta$ (``perfect signals'')\end{enumerate*}. The former corresponds with a Melitz economy with FE condition $\left(1-G\left(\varphi^{\ast}_{1}\right)\right)\bar{\pi}_{1}-\delta f_{e}=0$ and ZPC $\bar{\pi}_{1}=fk\left(\varphi^{\ast}_{1}\right)$, where, borrowing notation from \citet{melitzImpactTradeIntraIndustry2003}, $f_{e}\coloneqq f_{n}+f_{b}$; $G\left(\varphi\right)$ is the marginal distribution of $\varphi$; $k\left(\varphi^{\ast}\right)\coloneqq\left[\widetilde{\varphi}\left(\inf\Theta,\varphi^{\ast}\right)/\varphi^{\ast}\right]^{\sigma-1}-1$; and $\bar{\pi}\coloneqq\int_{\Theta}\widetilde{\pi}\left(\theta;\varphi^{\ast}\right)dC\left(\theta\right)$ are the \emph{ex ante} profit flows that can be expected from the entry process, net of entry costs. The ``perfect signals'' case, instead, leads to another Melitz economy with different parameters: the FE condition is $\left(1-G\left(\varphi^{\ast}_{2}\right)\right)\bar{\pi}_{2}-\delta f_{n}=0$, and the ZPC is $\bar{\pi}_{2}=\left(f+\delta{}f_{b}\right)k\left(\varphi^{\ast}_{2}\right)$.\footnote{Under independence, signals are uninformative and cannot prevent \emph{ex post} low-productivity entrants from paying $f_{b}$. With perfect signals, $\theta$ accurately predicts stage-three survival, hence the stage-two setup cost is isomorphic to an increase of the operational fixed cost by $\delta f_{b}$.}

The two cases are displayed on Panel A of Figure \ref{fig:LimitCases}, casting the relevant FE and ZPC curves on the $\left(\varphi^{\ast},\bar{\pi}\right)$ plane. Relative to the independence limit case, perfect signals yield sharper selection and higher productivity, but not necessarily higher average profits $\bar{\pi}$. If a planner attempted to address the relative inefficiency of independence through a tax on \emph{operational} labor costs (as in the comparative statics exercise from Figure \ref{fig:Equilibrium}) which is then rebated to consumers, it would shift the ZPC curve rightwards, but the FE curve would stay idle. As displayed on Panel B of Figure \ref{fig:LimitCases}, this could as well restore the productivity distribution obtained under perfect signals (thanks to tighter stage-three selection), but would invariably increase $\bar{\pi}$, thus depressing the equilibrium mass of firms/varieties $M$ and therefore, social welfare.\footnote{In equilibrium, labor $L$ is allocated between production employment ($L_{p}$) and the two entry activities. The equilibrium firm mass $M$ is increasing in $L_{p}$ and thus, $L$; it is also inversely related to average profits $\bar{\pi}$.}

\begin{figure}[ht!]
	\centering
	\caption{Equilibria and taxes with limit signals}
	\label{fig:LimitCases}
	\begin{subfigure}[h]{.5\textwidth}
    \centering
	\caption{The two limit cases}
	\begin{tikzpicture}
	    \begin{axis}
	        [axis lines = middle, samples = 100, domain=0:2, ymax = 2, x = 2.99cm, y = 2.6cm,
	        xlabel = $\varphi^{\ast}$, ylabel = $\bar{\pi}$, enlargelimits,
	        xtick={0.78906,1.22208}, xticklabels={$\varphi^{\ast}_{1}$,$\varphi^{\ast}_{2}$},
            x tick label style = {font = \small},
	        ytick={0.1,0.4}, yticklabels={$\delta{}f_{n}$,$\delta{}f_{e}$},
            y tick label style = {font = \small}]
	        \addplot [very thick] {0.6/x};
	        \addplot [very thick] {0.3*exp(x)+0.1};
	        \addplot [very thick, dashed] {1/x};
	        \addplot [very thick, dashed] {0.3*exp(x)-0.2};
	        \draw [densely dotted] (axis cs:0.78906,0) -- (axis cs:0.78906,0.760398);
	        \draw [densely dotted] (axis cs:1.22208,0) -- (axis cs:1.22208,0.818277);
	        \node [below] at (axis cs:0.4,0.81) {$\mathrm{FE}_{1}$};
	        \node [below] at (axis cs:0.4,0.50) {$\mathrm{FE}_{2}$};
	        \node [below] at (axis cs:1.8,0.55) {$\mathrm{ZPC}_{1}$};
	        \node [below] at (axis cs:1.8,0.80) {$\mathrm{ZPC}_{2}$};
	        \end{axis}
	    \end{tikzpicture}
	\end{subfigure}%
	\begin{subfigure}[h]{.5\textwidth}
    \centering
	\caption{Taxing operational wages}
	\begin{tikzpicture}
	    \begin{axis}
	        [axis lines = middle, samples = 100, domain=0:2, ymax = 2, x = 2.99cm, y = 2.6cm,
	        xlabel = $\varphi^{\ast}$, ylabel = $\bar{\pi}$, enlargelimits,
	        xtick={0.78906,1.22208}, xticklabels={$\varphi^{\ast}_{1}$,$\varphi^{\ast\ast}_{1}$},
            x tick label style = {font = \small},
	        ytick={0.4}, yticklabels={$\delta{}f_{e}$},
            y tick label style = {font = \small}]
	        \addplot [very thick] {0.6/x};
	        \addplot [very thick] {0.3*exp(x)+0.1};
	        \addplot [very thick, dashdotted] {1.36/x};
	        \addplot [draw=none] {1/x};
	        \addplot [draw=none] {0.3*exp(x)-0.2};
	        \draw [densely dotted] (axis cs:0.78906,0) -- (axis cs:0.78906,0.760398);
	        \draw [densely dotted] (axis cs:1.22208,0) -- (axis cs:1.22208,1.14542);
	        \node [below] at (axis cs:0.4,0.81) {$\mathrm{FE}_{1}$};
	        \node [below] at (axis cs:1.8,0.55) {$\mathrm{ZPC}_{1}$};
	        \node [below] at (axis cs:1.8,1.04) {$\mathrm{ZPC}_{1}^{\prime}$};
	    \end{axis}
	\end{tikzpicture}
	\end{subfigure}

    \note[Note]{Panel A: analysis of the two limit cases: independence (continuous lines) and perfect signals (dashed lines) as Melitz economies with different primitives. Panel B: the effect of taxing operational wages in the independence case; the dashed line is the new ZPC curve.}
\end{figure}

\section{Externalities}\label{sec:externalities}

We depart from the assumption of exogenous entry costs and examine instances where $f_{n}$ or $f_{b}$ (or both) depend on endogenous outcomes of the model. To simplify exposition and better illustrate the key insights, we restrict the analysis to two scenarios where both entry costs depend on the total mass of entrants $M_{e}$, or a subset of it. It is in a sense obvious that upon introducing externalities, the optimality result from Proposition \ref{prop:optimality} vanishes. By illustrating the incentives at work at their interaction, our analysis provides additional insights for fiscal policy. In particular, we show that slight variations between the two scenarios we entertain can lead to substantially different conclusions about the timing of fiscal policy intervention, and the potential for budget neutrality.

Before proceeding, some considerations are in order. First, upon endogenizing $f_{n}$ and $f_{b}$, equilibrium existence and uniqueness no longer follow from Proposition \ref{prop:benchmark}. Multiple equilibria are possible, with e.g. low $M_{e}$ or high $M_{e}$. The ensuing analysis is therefore \emph{local}: it applies around a particular equilibrium allocation, not necessarily unique, which is characterized by some mass of entrants $M_{e}$.\footnote{In what follows, we refer to \emph{interior} equilibria to rule out pathological (and uninteresting) cases where the model would predict zero or \emph{negative} entry $M_{e}$.} Second, we use $\tau_{n}\gtreqless0$ and $\tau_{b}\gtreqless0$ to denote taxes \emph{or subsidies} that a policymaker may implement at stage one or two of the entry game. These instruments deliver effective entry costs $f_{n}+\tau_{n}$ and $f_{b}+\tau_{b}$. We call pairs $(\tau_{n},\tau_{b})$ \emph{budget-neutral} if $\tau_{n}+[1-C\left(\theta^{\ast}\right)]\tau_{b}=0$: these can be implemented on a balanced budget without further fiscal manipulation. Such pairs are attractive as per \citet{dhingraMonopolisticCompetitionOptimum2019}, stage-three taxes and subsidies can introduce distortions. However, the results of this section are based on the solution of the constrained social planner problem, and can be theoretically implemented through lump-sum transfers or taxes on worker-consumers so as to keep a balanced fiscal budget.

\paragraph{Scenario 1: overall-entry externalities} In the first scenario, $f_{n}=f_{n}\left(M_{e}\right)$ and $f_{b}=f_{b}\left(M_{e}\right)$ are \emph{generic} functions of $M_{e}$. These can represent a variety of (positive) spillover and (negative) congestion effects that occur at either stage of the entry process. The straightforward implication of adding entry externalities is that in any decentralized equilibrium with entry $M_{e}$, the incentives of atomistic entrants diverge from the social planner's. At stage one, for example, the social cost of the marginal entrant is $f_{n}+M_{e}f_{n}^{\prime}\left(M_{e}\right)\gtreqless f_{n}$, and similarly at stage two. To express the resulting wedges succinctly, we denote the \emph{elasticities} of the entry costs $f_{n}\left(\cdot\right)$ and $f_{b}\left(\cdot\right)$ with respect to their arguments by $\varepsilon_{n}\left(\cdot\right)$ and $\varepsilon_{b}\left(\cdot\right)$, respectively. The welfare and fiscal policy implications of the first scenario are as follows.

\begin{prop}[Initial-entry externalities]\label{prop:breakdown} Let both entry costs be arbitrary functions of $M_{e}$. Any interior equilibrium is locally constrained-efficient if and only if:
\begin{equation*}
\tau_{n}^{\circ}:=\left[1-C\left(\theta^{\ast}\right)\right]\varepsilon_{b}\left(M_{e}\right)f_{b}\left(M_{e}\right)+\varepsilon_{n}\left(M_{e}\right)f_{n}\left(M_{e}\right)=0.
\end{equation*}
The efficient allocation can be implemented via Pigouvian taxes or subsidies $\left(\tau_{n},\tau_{b}\right)=\left(\tau_{n}^{\circ},0\right)$. If $\tau_{n}^{\circ}\neq0$, no budget-neutral tax instruments can restore efficiency.
\end{prop}

The intuition for this result is that in this scenario, the planner would ideally implement the following ``social'' version of the FE condition:
\begin{equation}\label{eq:social_free_entry}
\int_{\theta^{\ast}}^{\overline{\theta}}\frac{\widetilde{\pi}\left(\theta;\varphi^{\ast}\right)}{\delta}dC\left(\theta\right)-\left[1-C\left(\theta^{\ast}\right)\right]\left[1+\varepsilon_{b}\left(M_{e}\right)\right]f_{b}\left(M_{e}\right)-\left[1+\varepsilon_{n}\left(M_{e}\right)\right]f_{n}\left(M_{e}\right)=0.
\end{equation}
However, private entrants still comply with \eqref{eq:free_entry}: externalities introduce wedges on both entry costs whose magnitude depends on the respective elasticities at $M_{e}$. At the same time, once stage-one entry has occurred, the AC \eqref{eq:arbitrage} is socially efficient: intuitively, $f_{b}$ is already determined and is not further affected by stage-two decisions. To restore efficiency, the planner can introduce Pigouvian taxes or subsidies at stage one only, but since $\tau_{b}=0$, these instruments can never be budget-neutral. Intuitively, because low signals $\theta$ can induce potential entrants to exit at stage two, and thus altogether avoid engaging with the decision about paying $f_{b}$, $\tau_{n}^{\circ}$ is best implemented at stage one.

\paragraph{Scenario 2: staggered-selection externalities} The second scenario is only slightly different from the first: $f_{n}=f_{n}\left(M_{e}\right)$ but $f_{b}=f_{b}\left(M_{n}\right)$, where $M_{n}\coloneqq M_{e}\left[1-C\left(\theta^{\ast}\right)\right]$ is the expected mass of entrants passing stage-two selection in equilibrium. This is an arguably realistic scenario where only a restricted subset of entrants affect stage-two entry costs. Consider, for example, our microfoundation of stage-two selection based on financial intermediaries (``banks''). One can augment that microfoundation through channels that make it costlier for banks to screen an increasing number of business ideas. If low-signal entrants ($\theta<\theta^{\ast}$) self-select out of the screening process, they would not exert externalities on screening costs, although $f_{b}^{\prime}\left(M_{n}\right)>0$. This minor difference in the setup has relevant implications.

\begin{prop}[Staggered selection externalities]\label{prop:neutrality} Let $f_{n}$ and $f_{b}$ be arbitrary functions of $M_{e}$ and $M_{n}$, respectively. Any interior equilibrium is locally constrained-efficient if and only if:
\begin{align*}
\tau_{n}^{\ast}&:=\varepsilon_{n}\left(M_{e}\right)f_{n}\left(M_{e}\right)=0, \ \text{and:}\\
\tau_{b}^{\ast}&:=\varepsilon_{b}\left(M_{n}\right)f_{b}\left(M_{n}\right)=0.
\end{align*}
The efficient allocation can be implemented via Pigouvian taxes or subsidies $\left(\tau_{n},\tau_{b}\right)=\left(\tau_{n}^{\ast},\tau_{b}^{\ast}\right)$. These instruments are budget-neutral if $f_{n}^{\prime}\left(M_{e}\right)+\left[1-C\left(\theta^{\ast}\right)\right]^{2}f_{b}^{\prime}\left(M_{n}\right)=0$.
\end{prop}

This result differs from Proposition \ref{prop:breakdown} since the ``social'' version of the AC now is:
\begin{equation}\label{eq:social_arbitrage}
\frac{\widetilde{\pi}(\theta^{\ast};\varphi^{\ast})}{\delta}-\left[1+\varepsilon_{b}\left(M_{n}\right)\right]f_{b}\left(M_{n}\right)=0,
\end{equation}
because the signal threshold $\theta^{\ast}$ affects $M_{n}$ and therefore, $f_{b}$. Simultaneously, a variation of \eqref{eq:social_free_entry} applies in this scenario, though it involves $M_{n}$ (rather than $M_{e}$):
\begin{equation}\label{eq:social_free_entry_revisited}
\int_{\theta^{\ast}}^{\overline{\theta}}\frac{\widetilde{\pi}\left(\theta;\varphi^{\ast}\right)}{\delta}dC\left(\theta\right)-\left[1-C\left(\theta^{\ast}\right)\right]\left[1+\varepsilon_{b}\left(M_{n}\right)\right]f_{b}\left(M_{n}\right)-\left[1+\varepsilon_{n}\left(M_{e}\right)\right]f_{n}\left(M_{e}\right)=0.
\end{equation}
Therefore, a policymaker can restore efficiency through the stated Pigouvian instruments. These are \emph{coincidentally} budget-neutral if the two externalities affect entry costs in opposite directions (e.g., knowledge spillovers at stage one and screening congestion at stage two) \emph{and}, in equilibrium, the marginal effects on $f_{n}$ and $f_{b}$ are equal in magnitude up to the square of the stage-one survival probability. Consequently, the constant
\begin{equation}\label{eq:neutrality_measure}
    t^{\ast}\coloneqq\frac{1}{\left[1-C\left(\theta^{\ast}\right)\right]^{2}}\left|\frac{f_{n}^{\prime}\left(M_{e}\right)}{f_{b}^{\prime}\left(M_{n}\right)}\right|-1
\end{equation}
can be used to measure to what extent, in the target equilibrium, the efficient Pigouvian instruments deviate from the budget neutrality case, in which $t^{\ast}=0$.

\begin{figure}[ht!]
    \caption{Externalities, multiple equilibria, and budget neutrality}
    \label{fig:Externalities}
    \centering
    \makebox[\textwidth][c]{%
    \begin{tikzpicture}
        \begin{groupplot}[group style = {group size = 2 by 1, horizontal sep = 0.45cm},
                          axis lines = middle, samples = 200, scale only axis,
                          width = 6.9cm, height = 5.55cm, xmin = 0, xmax = 3.2,
                          xlabel = {$M_{e}$}, enlargelimits = false, clip = false]
        \nextgroupplot[ymin = 0, ymax = 0.85,
                       xtick = {0.280, 0.712, 1.119, 2.179, 2.747},
                       xticklabels = {$M_{e,P}^{1}$, $M_{e}^{1}$, $M_{e}^{\dagger}$, $M_{e,P}^{2}$,
                       $M_{e}^{2}$}, x tick label style = {font = \small},
                       ytick = \empty, ymajorticks = false, ylabel = {$\theta^{\ast}$}]
            \addplot [very thick] table [x = M, y = theta] {figures/figure3_private_ac.dat};
            \addplot [very thick, smooth] table [x = M, y = theta] {figures/figure3_private_fe.dat};
            \addplot [thick, densely dashed, black!55] table [x = M, y = theta] {figures/figure3_social_ac.dat};
            \addplot [thick, densely dashed, black!55, smooth] table [x = M, y = theta] {figures/figure3_social_fe.dat};
            \draw [densely dotted] (axis cs:0.712, 0) -- (axis cs:0.712, 0.313);
            \draw [densely dotted] (axis cs:2.747, 0) -- (axis cs:2.747, 0.516);
            \draw [densely dotted, gray] (axis cs:0.280, 0) -- (axis cs:0.280, 0.283);
            \draw [densely dotted, gray] (axis cs:2.179, 0) -- (axis cs:2.179, 0.594);
            \addplot [only marks, mark=*, mark size=2pt] coordinates { (0.712, 0.313) (2.747, 0.516) };
            \addplot [only marks, mark=diamond*, mark size=2.4pt,
                      mark options={draw=black!60, fill=white}, black!60]
                      coordinates { (0.280, 0.283) (2.179, 0.594) };
            \node [above left, font=\small] at (axis cs:0.9, 0.32) {$E_{1}$};
            \node [above left, font=\small] at (axis cs:2.9, 0.516) {$E_{2}$};
            \node [above right, font=\scriptsize, text=black!60] at (axis cs:0.01, 0.23) {$E_{1}^{P}$};
            \node [above left, font=\scriptsize, text=black!60] at (axis cs:2.3, 0.594) {$E_{2}^{P}$};
            \node [font=\footnotesize, fill=white, inner sep=1pt] at (axis cs:1.9, 0.45) {$\mathrm{AC}$};
            \node [font=\footnotesize, fill=white, inner sep=1pt] at (axis cs:1.6, 0.34) {$\mathrm{FE}$};
            \node [font=\scriptsize, text=black!60, anchor=west] at (axis cs:1.24, 0.76) {Dashed: \emph{social} AC and FE curves};
        \nextgroupplot[ymin = 0, ymax = 0.32, xtick = {0.280, 0.712, 1.119, 2.179, 2.747},
                       xticklabels = {$M_{e,P}^{1}$, $M_{e}^{1}$, $M_{e}^{\dagger}$, $M_{e,P}^{2}$, $M_{e}^{2}$}, x tick label style = {font = \small},
                       ytick = \empty, ymajorticks = false, ylabel = {$\chi$},
                       legend style = {at = {(0.80,0.90)}, anchor = north, draw = none}]
            \addplot [very thick] table [x = M, y = xi_n] {figures/figure3_budget_neutrality.dat};
            \addlegendentry{$\chi_{n}\left(M_{e}\right)$}
            \addplot [very thick, dashed] table [x = M, y = xi_b] {figures/figure3_budget_neutrality.dat};
            \addlegendentry{$\chi_{b}\left(M_{e}\right)$}
            \draw [densely dotted] (axis cs:1.119, 0) -- (axis cs:1.119, 0.078);
            \draw [densely dotted, gray] (axis cs:0.280, 0.125) -- (axis cs:0.280, 0.214);
            \draw [densely dotted, gray] (axis cs:0.712, 0.094) -- (axis cs:0.712, 0.128);
            \draw [densely dotted, gray] (axis cs:2.179, 0.022) -- (axis cs:2.179, 0.055);
            \draw [densely dotted, gray] (axis cs:2.747, 0.011) -- (axis cs:2.747, 0.047);
            \addplot [only marks, mark=*, mark size=2pt] coordinates { (1.119, 0.078) };
            \node [above right, font=\small, fill=white, inner sep=1pt] at (axis cs:1.119, 0.088) {$t^{\ast}=0$};
            \node [font=\scriptsize] at (axis cs:0.46, 0.135) {$t^{\ast}>0$};
            \node [font=\scriptsize] at (axis cs:2.50, 0.035) {$t^{\ast}<0$};
        \end{groupplot}
        \node [font=\small, anchor=south, yshift=0.36cm] at (group c1r1.north) {A. Decentralized and Pigouvian equilibria};
        \node [font=\small, anchor=south, yshift=0.36cm] at (group c2r1.north) {B. Deviation from budget neutrality};
    \end{tikzpicture}}%
    
    \note[Note]{Panel A: decentralized and Pigouvian equilibria, resulting from the intersection of the private or social AC and FE curves. Panel B: functions $\chi_{n}\left(M_{e}\right)$ and $\chi_{b}\left(M_{e}\right)$ as defined in the text, evaluated for all levels of entry $M_{e}$ traced in Panel A. All curves are based on a specific parametrization of the second scenario described in this section, where $C\left(\theta\right)$ is a standard uniform distribution, $G\left(\varphi\mid\theta\right)$ follows a Type-I Pareto distribution with scale parameter equal to $\theta$, $f_{b}\left(M_{n}\right)$ grows linearly in $M_{n}$, while $f_{n}\left(M_{e}\right)$ decays exponentially in $M_{e}$.}
\end{figure}

Budget neutrality and related welfare considerations must be assessed against potential equilibrium multiplicity. Figure \ref{fig:Externalities} illustrates this point using a particular parameterization of Scenario 2, summarized in the figure notes. Panel A displays the AC and FE conditions faced by firms, along with their ``social'' counterparts \eqref{eq:social_arbitrage} and \eqref{eq:social_free_entry_revisited}, on the $\left(M_{e},\theta^{\ast}\right)$ space. Whereas both AC curves are monotonically increasing due to congestion effects, the FE curves are U-shaped, as either congestion or spillover effects dominate for different values of $M_{e}$. Panel A displays two decentralized equilibria and two Pigouvian equilibria ($M_{e,P}^{1}$ and $M_{e,P}^{2}$); all are identified at the intersections between the relevant curves.\footnote{A specific productivity cutoff $\varphi^{\ast}$ is implicit in both equilibria (see the proofs of Propositions \ref{prop:breakdown} and \ref{prop:neutrality}).} Panel B evaluates budget neutrality for the same levels of entry $M_{e}$ traced in Panel A. It plots the two functions $\chi_{n}\left(M_{e}\right)\coloneqq\left|f_{n}^{\prime}\left(M_{e}\right)\right|$ and $\chi_{b}\left(M_{e}\right)\coloneqq[1-C\left(\theta^{\ast}\right)]^{2}f_{b}^{\prime}\left(M_{e}[1-C\left(\theta^{\ast}\right)]\right)$, where $\theta^{\ast}$ is the signal threshold identified by the AC for entry $M_{e}$. The intersection between these two curves, where $t^{\ast}=0$, occurs only at $M_{e}=M_{e}^{\dagger}$. That level coincides neither with any decentralized equilibrium nor, importantly, with either Pigouvian equilibrium.

Under the parameterization used to build Figure \ref{fig:Externalities}, one can calculate that the social welfare (inverse of the price level) associated with the low-entry Pigouvian equilibrium $M_{e,P}^{1}$ is approximately $0.091$. By contrast, for the high-entry Pigouvian equilibrium $M_{e,P}^{2}$, social welfare is approximately $0.574$: a more than sixfold increase. Neither equilibrium is budget-neutral, but the low-entry one requires more taxes than subsidies, and vice versa. Instances of equilibrium multiplicity such as this one provide a rationale for ``big push'' policies spurring entrepreneurship. It must be remarked, however, that this result follows from the particular functional form assumptions and parameterization we entertained to construct this example. Alternative choices can yield different scenarios, and different numbers of social equilibria. In general, our model is flexible enough to accommodate multiple scenarios (well beyond the two illustrated in this section) and policy implications.

\section{Calibration}\label{sec:calibration}

We perform an illustrative quantitative exercise based on the second scenario of Section \ref{sec:externalities}, aiming to decompose the mechanisms through which optimal Pigouvian instruments can raise welfare. In summary, this exercise is designed as follows. Assuming a realistic joint distribution of $\left(\theta,\varphi\right)$, we use U.S. data to calibrate a \emph{baseline} decentralized equilibrium. Using the AC and FE conditions, we solve for the resulting equilibrium values of $f_{n}$ and $f_{b}$. Given some specifications of the entry costs $f_{n}\left(M_{e}\right)$ and $f_{b}\left(M_{n}\right)$, we then solve for the social equilibrium that would result from \emph{local deviations} from the baseline equilibrium under multiple plausible hypotheses about the spillover and congestion elasticities $\varepsilon_{n}\left(M_{e}\right)$ and $\varepsilon_{b}\left(M_{n}\right)$. For each such hypothesis, we calculate the resulting change in the total mass of operating firms and in aggregate productivity: the two components of social welfare. We provide additional details and results in a dedicated appendix.

Our calibration leverages data on \emph{Business Applications} (BAs): filings for new Employer Identification Numbers recorded in the Census Bureau's \emph{Business Formation Statistics} (BFS). Selected BAs are classified as \emph{High-Propensity Business Applications} (HBAs), expected to yield Business Formations (operating companies) with a higher probability.\footnote{HBAs must meet one of four criteria: being a corporation, restructuring or acquiring an existing business, specifying a high-growth industry in the application, or mentioning a hiring date in it.} Those HBAs that also report a starting payroll date are further classified as \emph{Business Applications with Planned Wages} (WBAs). In addition, the BFS reports on those BAs that yield actual Business Formations four quarters after the application (BF4Q), as well as eight quarters (BF8Q). This layered information closely resembles the staged structure of our selection model. We conceptualize BAs moving to HBAs, and subsequently to the BF8Q (operational) stage, as the empirical counterpart of successful entrants of our model, which pass both thresholds $\theta^{\ast}$ and $\varphi^{\ast}$ following the experimentation stage. We think of WBAs as clearing a higher signal threshold $\theta^{w}>\theta^{\ast}$ which serves a statistical purpose unrelated to our model.\footnote{In the model, entrants failing to meet the signal threshold $\theta^{\ast}$ abandon the process, whereas in the data, non-HBA applications occasionally reach the business formation stage, though most typically at a later date. In this respect, the mapping between the BFS and our model must be considered approximate.}

We assume that $\left(\log\theta,\log\varphi\right)$ follows a bivariate normal distribution with correlation parameter $\rho>0$, whose marginals are $\log\theta\sim\mathcal{N}\left(0,1\right)$ and $\log\varphi\sim\mathcal{N}\left(\mu,1/\psi^{2}\right)$. We impose the normalization $\varphi^{\ast}=1$: thus, (log-)signals are measured on a normalized scale, whereas actual productivity is measured in units of the marginal entrant in the \emph{baseline} equilibrium. Under these choices, we calibrate key parameters of the model using selected empirical probabilities from the BFS, evaluated yearly, and averaged over the 2010-2021 cohorts. Let $\Phi\left(\cdot\right)$ be the cumulative standard normal distribution, and $\overline{\Phi}\left(\cdot,\cdot;\rho\right)$ the joint survival function of two standard normal random variables with correlation $\rho$. We set:
\begin{align*}
\Prob\left(\text{HBA}\mid\text{BA}\right)&=1-\Phi\left(\log\theta^{\ast}\right)=0.493\\
\Prob\left(\text{WBA}\mid\text{BA}\right)&=1-\Phi\left(\log\theta^{w}\right)=0.236
\end{align*}
hence, $\left(\theta^{\ast},\theta^{w}\right)=\left(1.018,2.053\right)$. Furthermore, since $\log\varphi^{\ast}=0$ in the baseline:
\begin{align*}
\Prob\left(\text{BF8Q}\cap\text{HBA}\mid\text{BA}\right)&=\overline{\Phi}\left(\log\theta^{\ast},-\mu\psi;\rho\right)=0.133\\
\Prob\left(\text{BF8Q}\cap\text{WBA}\mid\text{BA}\right)&=\overline{\Phi}\left(\log\theta^{w},-\mu\psi;\rho\right)=0.095
\end{align*}
hence, $\left(\rho,\mu\psi\right)=\left(0.644,-1.036\right)$. Separate identification of $\left(\mu,\psi\right)$ requires a fifth moment: it is natural to restrict the variance of log-productivity $\log\varphi>0$ among operating firms. Drawing on \citet{syverson2004product,syverson2011what} and \citet{foster2008reallocation},\footnote{As detailed in the appendix, we reconcile their dispersion measures with our model and with one another.} we set:
\begin{equation*}
\mathbb{V}\text{ar}\left[\log\varphi\mid\theta\geq\theta^{\ast},\varphi\geq1;\mu,\psi,\rho\right]=0.069.
\end{equation*}
By numerically solving the resulting integral we obtain $\left(\mu,\psi\right)=\left(-0.603,1.718\right)$.

We next characterize the baseline equilibrium; in what follows, we use the superscript $0$ to denote quantities evaluated specifically at that baseline. To operationalize the AC and FE conditions, values for $\sigma$ and $\delta$ are necessary. We set $\sigma=5$ (for a firm markup of 1.25) and $\delta=0.02$; the latter is a conversion of firm deaths as recorded by the Bureau's \emph{Business Dynamic Statistics}, divided by lagged firm counts, into a quarterly rate. We impose two more normalizations: $f=1$ and $M_{e}^{0}=1$ (the latter implies $M_{n}^{0}=0.493$); neither affects the ensuing welfare analysis. Thus, we solve for $f_{n}^{0}=46.390$ and $f_{b}^{0}=10.580$ by evaluating the AC and FE conditions at the baseline calibrated equilibrium. Lastly, we calculate total labor as the sum of total production costs and entry costs, obtaining $L=291.303$.

We examine the social optima associated with alternative, realistic hypotheses about the magnitude of both stage-one spillovers and stage-two negative externalities from the second scenario of Section \ref{sec:externalities}. We assume \emph{isoelastic} specifications of both entry costs:
\begin{equation*}
f_{n}\left(M_{e}\right)=f_{n}^{0}\left(\frac{M_{e}}{M_{e}^{0}}\right)^{\varepsilon_{n}^{0}}\quad\text{and}\quad f_{b}\left(M_{n}\right)=f_{b}^{0}\left(\frac{M_{n}}{M_{n}^{0}}\right)^{\varepsilon_{b}^{0}},
\end{equation*}
where $\varepsilon_{n}^{0}\leq0$ and $\varepsilon_{b}^{0}\geq0$ encode spillover and congestion forces, respectively. The literature offers no direct guidance about these parameters; however, positive entry spillovers are conceptually (though vaguely) related to spillover and agglomeration effects in urban and innovation economics. Our baseline welfare analysis is based on $\left|\varepsilon_{n}^{0}\right|\in\left\{0.03,0.05,0.08\right\}$: three reference values inspired by that literature's elasticity estimates for productivity and wages \citep{combes2008spatial,combes2015empirics}.\footnote{Specifically, the survey by \citet{combes2015empirics} reports estimates clustered in the $\left[0.02,0.10\right]$ range, with sorting-corrected estimates such as \citet{combes2008spatial} at the lower end.} For each of these three values, we evaluate congestion elasticities over the range $\varepsilon_{b}^{0}\in\left[0.0,0.1\right]$.

Guided by Proposition \ref{prop:neutrality}, for every pair $\left(\varepsilon_{n}^{0},\varepsilon_{b}^{0}\right)$ that we examine we solve for the triple $\left(M_{e},\theta^{\ast},\varphi^{\ast}\right)$ that satisfies both ``social'' AC and FE conditions \eqref{eq:social_arbitrage} and \eqref{eq:social_free_entry_revisited} along with the (binding) resource constraint.\footnote{We search for numerical solutions using the baseline calibrated equilibrium as the starting value. The resource constraint keeps $L=291.303$ constant and sets it as the sum of total production and entry costs.} In the resulting social optima $\varphi^{\ast}$ may differ from one. We thus calculate the steady-state mass of firms $M=\overline{\Phi}\left(\log\theta^{\ast},\psi\left(\log\varphi^{\ast}-\mu\right);\rho\right) M_{e}/\delta$ and aggregate productivity $\widetilde{\varphi}\left(\theta^{\ast},\varphi^{\ast}\right)$, and we evaluate their change relative to the calibrated baseline: the results are displayed in Panels A and B of Figure \ref{fig:calibration_exercise}, respectively. In our central scenario where $-\varepsilon_{n}^{0}=\varepsilon_{b}^{0}=0.05$, $M$ increases by about 0.25 percent (as stronger entry incentives dominate tighter selection at stages two and three) whereas $\widetilde{\varphi}\left(\theta^{\ast},\varphi^{\ast}\right)$ increases by about 0.75 percent (since selection acts on a larger entrant pool). As spillover effects become stronger ($\varepsilon_{n}^{0}$ decreases) both variables rise more markedly, and vice versa. By contrast, as congestion rises ($\varepsilon_{b}^{0}$ moves away from zero) selection tightens at stage two, with markedly negative effects on $M$ (which can even fall relative to the baseline) and positive but mild effects on productivity. As we show in the appendix, the results are qualitatively similar whether one defines successful entry as BF4Q or perturbs $\sigma$.

\begin{figure}[ht!]
    \centering
    \caption{Welfare determinants in the calibrated social optima}
    \label{fig:calibration_exercise}
    \begin{tikzpicture}
    \begin{groupplot}[
        group style={group size=2 by 1, horizontal sep=1cm},
        axis lines=left,
        axis line style={->},
        scale only axis,
        width=0.405\textwidth,
        height=4.85cm,
        xmin=0,
        xmax=0.100,
        enlarge x limits={upper},
        enlarge y limits={upper},
        xtick={0,0.02,0.04,0.06,0.08,0.10},
        xticklabels={0.00,0.02,0.04,0.06,0.08,0.10},
        scaled x ticks=false,
        xlabel={Congestion elasticity $\varepsilon_{b}^{0}$},
        ylabel={Percentage change},
        label style={font=\small},
        tick label style={font=\small},
        legend style={font=\small, fill=white, draw=black!35, rounded corners=0pt, inner sep=2pt, /tikz/every even column/.append style={column sep=0.35cm}},
        clip=true]
        \nextgroupplot[
            title={A. Operating firm mass},
            title style={font=\small},
            ymin=-0.75,
            ymax=1.25,
            ytick={-0.5,0,0.5,1.0},
            legend to name=figurefourlegend,
            legend columns=3]
            \addplot[very thick, black!55, densely dotted]
                table[x=epsilon_b,y=firmmass_eps003, col sep=comma] {figures/figure4_panel_a.csv};
            \addlegendentry{$\varepsilon_n^0=-0.03$}
            \addplot[very thick, black]
                table[x=epsilon_b,y=firmmass_eps005, col sep=comma] {figures/figure4_panel_a.csv};
            \addlegendentry{$\varepsilon_n^0=-0.05$}
            \addplot[very thick, black!70, dashed]
                table[x=epsilon_b,y=firmmass_eps008, col sep=comma] {figures/figure4_panel_a.csv};
            \addlegendentry{$\varepsilon_n^0=-0.08$}
        \nextgroupplot[
            title={B. Aggregate productivity},
            title style={font=\small},
            ymin=0,
            ymax=1.4,
            ytick={0,0.4,0.8,1.2},
            ylabel={}]
            \addplot[very thick, black!55, densely dotted]
                table[x=epsilon_b,y=productivity_eps003, col sep=comma] {figures/figure4_panel_b.csv};
            \addplot[very thick, black]
                table[x=epsilon_b,y=productivity_eps005, col sep=comma] {figures/figure4_panel_b.csv};
            \addplot[very thick, black!70, dashed]
                table[x=epsilon_b,y=productivity_eps008, col sep=comma] {figures/figure4_panel_b.csv};
    \end{groupplot}
    \node[anchor=north, yshift=-1.35cm] at ($(group c1r1.south east)!0.5!(group c2r1.south west)$) {\pgfplotslegendfromname{figurefourlegend}};
    \end{tikzpicture}
    \note[Note]{Across social optima identified by a pair $\smash{\left(\varepsilon_{n}^{0},\varepsilon_{b}^{0}\right)}$, this figure reports percentage changes, relative to the calibrated decentralized baseline, of firm mass $M$ (Panel A) and aggregate productivity $\smash{\widetilde{\varphi}\left(\theta^{\ast},\varphi^{\ast}\right)}$ (Panel B). Each curve corresponds to a different value of the spillover elasticity $\smash{\left|\varepsilon_{n}^{0}\right|}\in\left\{0.03,0.05,0.08\right\}$. The horizontal axes run through the range $\smash{\varepsilon_{b}^{0}}\in\left[0.0,0.1\right]$ for the congestion elasticity. Throughout, $\sigma=5$ holds.}
\end{figure}

From these results one can extrapolate the overall welfare effects. In logarithms:
\begin{equation*}
\log U=\log\left(\frac{\sigma-1}{\sigma}\right)+\frac{1}{\sigma-1}\log M+\log\widetilde{\varphi}\left(\theta^{\ast},\varphi^{\ast}\right).
\end{equation*}
Hence, by combining the results of the two panels in Figure \ref{fig:calibration_exercise}, we evaluate welfare gains between 0.4 and 1.5 percent. These moderate figures deserve a cautious interpretation. First, they hinge on conservative spillover elasticities and assumptions maintained across all firms and industries, such as regular substitution patterns and identical entry costs. Second, they assume no welfare distortions from fiscal policy: as we show in the appendix, the optimal Pigouvian instruments typically come with a positive net cost, ideally financed via non-distortionary means (e.g., lump-sum taxes on worker-consumers). Third, they are based on local searches around the baseline, thus ignoring the potential of stronger gains under equilibrium multiplicity, as in the ``big push'' example of Figure \ref{fig:Externalities}.

\section{Conclusion}\label{sec:conclusion}

We develop a theory of staged firm entry featuring incremental acquisition of information about future performance, and externalities on entry costs. Its key policy implication is that externalities may be offset by welfare-enhancing Pigouvian taxes and subsidies if these are timed at the key firm selection gates. The analysis and results presented in this paper are based on a benchmark setup with CES preferences, which feature no post-entry inefficiencies. The next step is to analyze variations of the model departing from this benchmark, and characterize the resulting optimal fiscal policy. A more realistic model can motivate a more nuanced empirical quantification of the externalities at play, and the welfare gains from optimal policy. We leave these undertakings to future work.

\bibliography{\bib}

\newpage
\appendix
\singlespacing
\footnotesize

\section*{Mathematical proofs}

\begin{proof}[Proof of Proposition \ref{prop:benchmark}] This proof studies the Arbitrage and Free Entry conditions, and their intersection. Before proceeding, it is useful to observe that Assumptions \ref{ass:A2} and \ref{ass:A3}, combined, ensure uniform integrability of the family of distributions $\smash{\left\{\varphi^{\sigma-1}\ \text{under}\ G\left(\cdot\mid\theta\right)\right\}_{\theta\in\left(\underline{\theta},\theta_{c}\right]}}$, where $\theta_{c}\in\Theta$. In fact, for any $K\in\mathbb{R}_{++}$:
\begin{equation*}
\mathbb{E}\left[\varphi^{\sigma-1}\mathds{1}\left\{\varphi^{\sigma-1}>K\right\}\mid\theta\right]\leq\mathbb{E}\left[\varphi^{\sigma-1}\mathds{1}\left\{\varphi^{\sigma-1}>K\right\}\mid\theta_{c}\right]
\end{equation*}
as $G\left(\cdot\mid\theta_{c}\right)$ first-order stochastically dominates $G\left(\cdot\mid\theta\right)$; and the right-hand side of the inequality vanishes as $K\to\infty$ because $\mathbb{E}\left[\varphi^{\sigma-1}\mid\theta_{c}\right]<\infty$. By Assumption \ref{ass:A1} and Vitali's theorem, this yields continuity of $\widetilde{\pi}\left(\theta;\varphi^{\ast}\right)$ with respect to $\theta$.

We begin by analyzing the Arbitrage Condition $\widetilde{\pi}\left(\theta^{\ast};\varphi^{\ast}\right)=\delta f_{b}$. We first note that $\smash{\widetilde{\pi}\left(\theta;\varphi^{\ast}\right)}$ is strictly decreasing in $\varphi^{\ast}$, since higher productivity thresholds depress stage-two expected profits. Then, fix $\varphi^{\ast}>0$. Note that by Assumption \ref{ass:A4}, the above uniform integrability result, and Vitali's Theorem:
\begin{equation*}
\widetilde{\pi}\left(\theta;\varphi^{\ast}\right) \to 0 \quad\text{as }\theta\downarrow\underline{\theta},
\end{equation*}
because at the limit, $\Prob\left([\varphi^{\ast},\infty)\mid\theta\right)\to 0$. Conversely, for every $\varphi>\varphi^{\ast}$:
\begin{equation*}
\widetilde{\pi}\left(\theta;\varphi^{\ast}\right) \geq f\left[\left(\frac{\varphi}{\varphi^{\ast}}\right)^{\sigma-1}-1\right]\Prob\left(\left[\varphi,\infty\right)\mid\theta\right),
\end{equation*}
and since $\varphi$ can be chosen arbitrarily large, it follows that at the limit where $\Prob\left([\varphi,\infty)\mid\theta\right)\to 1$:
\begin{equation*}
\widetilde{\pi}\left(\theta;\varphi^{\ast}\right) \to \infty \quad\text{as }\theta\uparrow\overline{\theta}.
\end{equation*}
Therefore, by continuity and strict monotonicity of $\widetilde{\pi}\left(\theta;\varphi^{\ast}\right)$, for every $\varphi^{\ast}>0$ there exists a unique $\theta^{\ast}\!\left(\varphi^{\ast}\right)\in\Theta$ such that the Arbitrage Condition holds:
\begin{equation*}
\widetilde{\pi}\left(\theta^{\ast}\!\left(\varphi^{\ast}\right);\varphi^{\ast}\right)=\delta f_{b}.
\end{equation*}
Thus, the Arbitrage Condition defines a unique map $\varphi^{\ast}\mapsto\theta^{\ast}\!\left(\varphi^{\ast}\right)$. This map is strictly increasing because $\smash{\widetilde{\pi}\left(\theta;\varphi^{\ast}\right)}$ is strictly decreasing in $\varphi^{\ast}$ and Assumption \ref{ass:A3} (to motivate stage-two costs when $\varphi^{\ast}$ rises, $\theta^{\ast}$ must rise too). Note that for $\theta>\theta^{\ast}\!\left(\varphi^{\ast}\right)$ it is $\smash{\widetilde{\pi}\left(\theta;\varphi^{\ast}\right)>\delta f_{b}}$, and vice versa.

We next analyze the Free Entry condition, which we recast as $\smash{\gamma\left(\theta^{\ast},\varphi^{\ast}\right)=0}$, where:
\begin{equation*}
\gamma\left(\theta^{\ast},\varphi^{\ast}\right)\coloneqq\int_{\theta^\ast}^{\overline{\theta}}\left[\frac{\widetilde{\pi}\left(\theta;\varphi^{\ast}\right)}{\delta}-f_{b}\right]dC\left(\theta\right)-f_{n}.
\end{equation*}
One can easily verify that this function is decreasing in $\varphi^{\ast}$, and that given $\varphi^{\ast}$, its derivative with respect to $\theta^{\ast}$ changes sign at $\theta^{\ast}\!\left(\varphi^{\ast}\right)$. Hence:
\begin{equation*}
\arg\max_{\theta\in\Theta}\gamma\left(\theta,\varphi^{\ast}\right)=\theta^{\ast}\!\left(\varphi^{\ast}\right).
\end{equation*}
Together with the Implicit Function Theorem, these observations support the Proposition's claim that in equilibrium, the AC and FE curves cross at the maximum of the FE implicit function. Next, define:
\begin{equation*}
\widetilde{\gamma}\left(\varphi^{\ast}\right)\coloneqq\gamma\left(\theta^{\ast}\!\left(\varphi^{\ast}\right),\varphi^{\ast}\right).
\end{equation*}
By the earlier uniform integrability result, this function inherits continuity from $\widetilde{\pi}(\cdot)$ and $\gamma(\cdot)$. Furthermore, it is decreasing in $\varphi^{\ast}$. To see this, take $\varphi_{2}^{\ast}>\varphi_{1}^{\ast}$. Since $\widetilde{\pi}\left(\theta,\varphi^{\ast}\right)$ is strictly decreasing in $\varphi^{\ast}$, for every $\theta\in\Theta$:
\begin{equation*}
\gamma\left(\theta,\varphi_{2}^{\ast}\right)<\gamma\left(\theta,\varphi_{1}^{\ast}\right).    
\end{equation*}
Taking maxima of the above with respect to $\theta$ yields $\smash{\widetilde{\gamma}\left(\varphi_{2}^{\ast}\right)<\widetilde{\gamma}\left(\varphi_{1}^{\ast}\right)}$.

To complete the proof, it is necessary to show that equation $\smash{\widetilde{\gamma}\left(\varphi^{\ast}\right)=0}$ has a unique root $\varphi^{\ast}$, where the Arbitrage Condition and the Free Entry condition are simultaneously satisfied. To this end, we examine the behavior of the function as $\varphi^{\ast}\downarrow0$ and $\varphi^{\ast}\uparrow\infty$. First, consider $\varphi^\ast\downarrow 0$ and fix any $\varphi>0$. By Assumption \ref{ass:A4}, there exists $\theta_{0}\in\Theta$ such that for all $\theta\geq\theta_{0}$,
\begin{equation*}
\Prob\left(\left[\varphi,\infty\right)\mid\theta\right)\geq\frac{1}{2}.
\end{equation*}
Then for all such $\theta\geq\theta_{0}$,
\begin{equation*}
\widetilde{\pi}\left(\theta,\varphi^{\ast}\right)\geq\frac{f}{2}\left[\left(\frac{\varphi}{\varphi^\ast}\right)^{\sigma-1}-1\right].
\end{equation*}
Therefore:
\begin{equation*}
\gamma\left(\theta_{0},\varphi^{\ast}\right)\geq\left[1-C\left(\theta_{0}\right)\right]
\left\{\frac{f}{2\delta}\left[\left(\frac{\varphi}{\varphi^{\ast}}\right)^{\sigma-1}-1\right]-f_{b}\right\}-f_{n}.
\end{equation*}
As $\varphi^{\ast}\downarrow0$, the right-hand side diverges to $+\infty$. Hence:
\begin{equation*}
\widetilde{\gamma}\left(\varphi^{\ast}\right)\to +\infty\qquad\text{as }\varphi^\ast\downarrow0.
\end{equation*}
Now consider $\varphi^\ast\uparrow\infty$. Observe that, given the definition of $\widetilde{\pi}(\cdot)$ in terms of a truncated moment of $\varphi$, it is:
\begin{equation*}
\widetilde{\pi}\left(\theta,\varphi^{\ast}\right)\leq f\frac{\mathbb{E}\left[\varphi^{\sigma-1}\mid \theta\right]}{\left(\varphi^{\ast}\right)^{\sigma-1}}.
\end{equation*}
Therefore, for every $\theta\in\Theta$,
\begin{equation*}
\gamma\left(\theta,\varphi^{\ast}\right)+f_{n}\leq\int_{\theta^{\ast}}^{\overline{\theta}}\frac{\widetilde{\pi}\left(\theta,\varphi^{\ast}\right)}{\delta} dC\left(\theta\right)
\leq\frac{f}{\delta}(\varphi^\ast)^{-\left(\sigma-1\right)}\int_{\theta^{\ast}}^{\overline{\theta}}
\mathbb{E}\left[\varphi^{\sigma-1}\mid\theta\right]dC\left(\theta\right)\leq
\frac{f}{\delta}\frac{\mathbb{E}\left[\varphi^{\sigma-1}\right]}{\left(\varphi^{\ast}\right)^{\sigma-1}}.
\end{equation*}
Taking the maximum over $\theta$ yields:
\begin{equation*}
\widetilde{\gamma}\left(\varphi^{\ast}\right)+f_{n}\leq\frac{f}{\delta}\frac{\mathbb{E}\left[\varphi^{\sigma-1}\right]}{\left(\varphi^{\ast}\right)^{\sigma-1}}.
\end{equation*}
By Assumption \ref{ass:A2}, the right-hand side converges to zero as $\varphi^\ast\uparrow\infty$. Hence:
\begin{equation*}
\widetilde{\gamma}\left(\varphi^{\ast}\right)\to-f_{n}\qquad\text{as }\varphi^{\ast}\uparrow\infty.
\end{equation*}
Because $\widetilde{\gamma}\left(\varphi^{\ast}\right)$ is continuous and strictly decreasing, its limit behavior implies the existence of a unique interior root. This completes the proof.
\end{proof}

\begin{proof}[Proof of Proposition \ref{prop:optimality}] This proof is adapted from \citet{dhingraMonopolisticCompetitionOptimum2019}. We consider a planner that maximizes the utility of the representative consumer, i.e., the following CES aggregator:
\begin{equation*}
U^{\frac{\sigma-1}{\sigma}}=\int_{\omega\in\Omega}q\left(\omega\right)^\frac{\sigma-1}{\sigma}d\omega    
\end{equation*}
subject to the economy's labor resource constraint. Let $\lambda>0$ denote the shadow value of one unit of labor. The planner solves the problem recursively, starting from stage three assuming that entry has occurred, and proceeding backwards to the two initial entry stages. At every stage, the planner has the same information as agents (entrants/firms) in the decentralized economy.

We start with stage three: we fix $\lambda$, and we focus on a specific variety $\omega$ for which both entry costs have been paid. Having observed $\varphi\left(\omega\right)$, the planner chooses whether to operate and, if so, how much to produce:
\begin{equation*}
\max_{q\geq0}\left\{q^{\frac{\sigma-1}{\sigma}}-\lambda\left(f\cdot\mathds{1}\left[q>0\right]+\frac{q}{\varphi}\right)\right\},  
\end{equation*}
where for simplicity dependence on $\omega$ is removed from $q\left(\omega\right)$ and $\varphi\left(\omega\right)$. If production is strictly positive, the first-order condition yields the solution:
\begin{equation*}
q\left(\varphi\right)=\left(\frac{\sigma-1}{\sigma}\frac{\varphi}{\lambda}\right)^{\sigma}.  
\end{equation*}
Therefore, the planner's quantity schedule is a power function of productivity with exponent $\sigma$, exactly as in the conventional decentralized equilibrium. The \emph{flow} social value (surplus) from producing $q(\omega)$ in a financed firm with productivity $\varphi(\omega)$ is:
\begin{equation*}
\pi^{P}(\varphi;\lambda)=\max\left\{\frac{1}{\sigma}\left(\frac{\sigma-1}{\sigma}\frac{\varphi}{\lambda}\right)^{\sigma-1}-\lambda f,0\right\}.
\end{equation*}
Following standard manipulation, this expression can be written in terms of the cutoff productivity $\varphi^{\ast}$ that sets $\pi^{P}(\varphi^{\ast};\lambda)=0$:
\begin{equation*}
\pi^{P}\left(\varphi;\varphi^{\ast},\lambda\right)=\max\left\{\lambda f\left[\left(\frac{\varphi}{\varphi^{\ast}}\right)^{\sigma-1}-1\right],0\right\}.
\end{equation*}
Production is thus socially valuable if and only if $\varphi>\varphi^{\ast}$ and $\pi^{P}\left(\varphi;\varphi^{\ast},\lambda\right)>0$. Further inspection reveals that $\varphi^{\ast}$ coincides with the decentralized productivity cutoff, since for $\varphi>\varphi^{\ast}$, private profits are:
\begin{equation*}
\pi\left(\varphi;\varphi^{\ast}\right)=f\left[\left(\frac{\varphi}{\varphi^{\ast}}\right)^{\sigma-1}-1\right].
\end{equation*}
Consequently:
\begin{equation*}
\pi^{P}\left(\varphi;\varphi^{\ast},\lambda\right)=\lambda\,\pi\left(\varphi;\varphi^{\ast}\right),
\end{equation*}
that is, the social value of production equals private profits multiplied by the proportionality factor $\lambda$.

We now move back to stage two. After observing $\theta$ but before observing $\varphi$, the planner decides whether to incur the financing cost $f_{b}$ for each variety $\omega$. Let
\begin{equation*}
\widetilde{\pi}^{P}\left(\theta;\varphi^{\ast},\lambda\right)\coloneqq\mathbb{E}_{\varphi\mid\theta}\!\left[\pi^{P}\left(\varphi;\varphi^{\ast},\lambda\right)\mid \theta\right]
\end{equation*}
denote the planner's \emph{expected} flow surplus from financing a variety with signal $\theta$. Using the step three results,
\begin{equation*}
\widetilde{\pi}^{P}\left(\theta;\varphi^{\ast},\lambda\right)=\lambda\,\mathbb{E}_{\varphi\mid\theta}\!\left[\pi\left(\varphi;\varphi^{\ast}\right)\mid\theta\right]=\lambda\,\widetilde{\pi}\left(\theta;\varphi^{\ast}\right),
\end{equation*}
where $\widetilde{\pi}\left(\theta;\varphi^{\ast}\right)$ is the expected flow private profit defined in \eqref{eq:pi-tilde}. The planner finances a variety if and only if its expected present value exceeds the setup cost $\lambda f_{b}$:
\begin{equation*}
\frac{\widetilde{\pi}^{P}\left(\theta;\varphi^{\ast},\lambda\right)}{\delta}=\frac{\lambda\widetilde{\pi}\left(\theta;\varphi^{\ast}\right)}{\delta}\geq\lambda f_{b}.
\end{equation*}
By Assumption \ref{ass:A3}, this problem admits a threshold signal $\theta^{\ast}\in\Theta$ satisfying $\widetilde{\pi}^{P}\left(\theta^{\ast};\varphi^{\ast},\lambda\right)=\lambda\widetilde{\pi}\left(\theta^{\ast};\varphi^{\ast}\right)=\delta\lambda f_{b}.$ This threshold is clearly the same as the one satisfying the Arbitrage Condition \eqref{eq:arbitrage}.

Lastly, we examine the planner problem at stage one. Before observing the signal, the planner evaluates the social value of one additional variety. For every additional entry experiment, society pays $f_n$ for sure, pays $f_b$ only if $\theta\geq\theta^{\ast}$, and in that event receives the discounted continuation surplus associated with financing. Thus, the planner's net value of one more experiment is
\begin{equation*}
\gamma^{P}\left(\theta^{\ast},\varphi^{\ast};\lambda\right)=\int_{\theta^{\ast}}^{\overline{\theta}}\frac{\widetilde{\pi}^{P}\left(\theta;\varphi^{\ast},\lambda\right)}{\delta}\,dC\left(\theta\right)
-\lambda\left[1-C\left(\theta^{\ast}\right)\right]f_{b}-\lambda f_{n}.
\end{equation*}
Using again $\pi^{P}\left(\varphi;\lambda\right)=\lambda\,\pi\left(\varphi\right)$, it is:
\begin{equation*}
\gamma^{P}\left(\theta^{\ast},\varphi^{\ast};\lambda\right)=\lambda\gamma\left(\theta^{\ast},\varphi^{\ast}\right)=\lambda\left[\int_{\theta^\ast}^{\overline{\theta}}\left[\frac{\widetilde{\pi}\left(\theta;\varphi^{\ast}\right)}{\delta}-f_{b}\right]dC\left(\theta\right)-f_{n}\right].
\end{equation*}
At an interior optimum, the marginal experiment must yield zero net value, i.e.:
\begin{equation*}
\gamma^{P}\left(\theta^{\ast},\varphi^{\ast};\lambda\right)=0.
\end{equation*}
Since $\lambda>0$, this is equivalent to the Free Entry condition in \eqref{eq:free_entry}.

In summary, the planner's optimality conditions coincide with the decentralized equilibrium conditions: the planner chooses the same quantity schedule $q\left(\varphi\right)$, the same productivity threshold $\varphi^{\ast}$, the same signal threshold $\theta^{\ast}$ and the same experimentation cutoff rule as decentralized agents. Hence, every decentralized equilibrium allocation solves the planner's constrained problem. As the equilibrium pair $(\theta^{\ast},\varphi^{\ast})$ is unique by Proposition \ref{prop:benchmark}, the decentralized equilibrium is the unique constrained-efficient allocation.
\end{proof}

\begin{proof}[Proof of Proposition \ref{prop:breakdown}] We formulate the planner problem under the illustrative externalities from the text. The problem is again best formulated recursively, starting from stage three and proceeding backwards. At stage three, the planner takes entry as given and observes the productivity $\varphi$ of all firms. Given the shadow value of labor $\lambda$, the planner maximizes $q^{\smash{\nicefrac{\sigma-1}{\sigma}}}-\lambda\left(f+q/\varphi\right)$ over $q$ for each firm, leading to the same solution as in the benchmark planner problem from Proposition \ref{prop:optimality} and in the decentralized equilibrium. Thus, any source of inefficiency must stem from the earlier experimentation and financing stages.

Due to potential externalities, the planner cannot evaluate the value of each experimentation in isolation: thus, it maximizes the representative consumer's utility $U^{\smash{\nicefrac{\sigma-1}{\sigma}}}$ under the resource (labor) constraint. Let:
\begin{equation*}
p\left(\theta^{\ast},\varphi^{\ast}\right)\coloneqq\Prob\left(\theta\geq\theta^{\ast},\varphi\geq\varphi^{\ast}\right)=\int_{\theta^{\ast}}^{\overline{\theta}}\left[1-G\left(\varphi^{\ast}\mid\theta\right)\right]dC\left(\theta\right).
\end{equation*}
Leveraging the \emph{steady-state identity} $\delta M = M_{e}p\left(\theta^{\ast},\varphi^{\ast}\right)$, the resource constraint can be expressed as:
\begin{equation*}
L\geq M_{e}\left[\left(\sigma-1\right)\frac{f}{\delta}\int_{\theta^{\ast}}^{\overline{\theta}}\int_{\varphi^{\ast}}^{\infty}\left(\frac{\varphi}{\varphi^{\ast}}\right)^{\sigma-1}\,dG\left(\varphi\mid\theta\right)\,dC\left(\theta\right)+\frac{f}{\delta}p\left(\theta^{\ast},\varphi^{\ast}\right)+\left[1-C\left(\theta^{\ast}\right)\right]f_{b}\left(M_{e}\right)+f_{n}\left(M_{e}\right)\right].
\end{equation*}
In each period, total labor $L$ must be allocated to variable production costs (for each firm, these are equal to their revenues $r(\varphi)$ divided by their markup); fixed production costs $Mf$; and the two entry costs, which themselves depend on $M_{e}$. Our focus on interior equilibria adds another constraint: $M_{e}\geq0$. The treatment of this simple inequality constraint via Kuhn-Tucker conditions is trivial, and is therefore omitted. The ensuing analysis is understood as a constrained maximization problem that can potentially yield multiple solutions, among which those featuring $M_{e}\leq0$ are uninteresting. The focus lies on the optimality conditions.

The planner may control the productivity cutoff $\varphi^{\ast}$, the signal cutoff $\theta^{\ast}$, and the mass of entrants $M_{e}$. However, given $\theta^{\ast}$ and $M_{e}$, the productivity threshold $\varphi^{\ast}$ is determined at stage three of the model as a sole function of fixed parameters and the shadow value of labor $\lambda$. The proof of Proposition \ref{prop:optimality} in fact shows that:
\begin{equation*}
\left(\frac{\sigma-1}{\sigma}\frac{\varphi^{\ast}}{\lambda}\right)^{\sigma-1}=\lambda\sigma f,
\end{equation*}
which follows from $\smash{\pi^{P}\left(\varphi^{\ast};\lambda\right)=0}$. It turns out that by anticipating the stage-three optimal production choices, one can formulate a more straightforward planner problem in terms of $\left(\theta^{\ast},M_{e},\lambda\right)$. Observe that:
\begin{align*}
U^{\frac{\sigma-1}{\sigma}}&=\frac{M}{p\left(\theta^{\ast},\varphi^{\ast}\right)}\int_{\theta^{\ast}}^{\overline{\theta}}\int_{\varphi^{\ast}}^{\infty}\left(q\left(\varphi\right)\right)^{\frac{\sigma-1}{\sigma}}dG\left(\varphi\mid\theta\right)\,dC\left(\theta\right)\\
&=\frac{M_{e}}{\delta}\left(\frac{\sigma-1}{\sigma}\right)^{\sigma-1}\int_{\theta^{\ast}}^{\overline{\theta}}\int_{\varphi^{\ast}}^{\infty}\left(\frac{\varphi}{\lambda}\right)^{\sigma-1}dG\left(\varphi\mid\theta\right)\,dC\left(\theta\right)\\
&=\frac{\lambda M_{e}\sigma f}{\delta}\int_{\theta^{\ast}}^{\overline{\theta}}\int_{\varphi^{\ast}}^{\infty}\left(\frac{\varphi}{\varphi^{\ast}}\right)^{\sigma-1}dG\left(\varphi\mid\theta\right)\,dC\left(\theta\right)
\end{align*}
where the first equality applies the definition of $U$, the second equality follows from the expression of optimal quantities at stage three and the steady-state identity, whereas the third equality leverages the stage-three relationship $\smash{\pi^{P}\left(\varphi^{\ast};\lambda\right)=0}$ linking $\varphi^{\ast}$ with $\lambda$, $\sigma$ and $f$. The Lagrangian of the \emph{reduced planner problem} is thus:
\begin{equation*}
\mathcal{L}=\lambda\left\{L+M_{e}\left[\frac{f}{\delta}\int_{\theta^{\ast}}^{\overline{\theta}}\int_{\varphi^{\ast}}^{\infty}\left(\frac{\varphi}{\varphi^{\ast}}\right)^{\sigma-1}\,dG\left(\varphi\mid\theta\right)\,dC\left(\theta\right)-\frac{f}{\delta}p\left(\theta^{\ast},\varphi^{\ast}\right)-\left[1-C\left(\theta^{\ast}\right)\right]f_{b}\left(M_{e}\right)-f_{n}\left(M_{e}\right)\right]\right\}.
\end{equation*}
This Lagrangian is recast even more succinctly using \eqref{eq:pi-tilde} and the integral representation of $p\left(\theta^{\ast},\varphi^{\ast}\right)$:
\begin{equation*}
\mathcal{L}\left(\theta^{\ast},M_{e},\lambda\right)=\lambda\left[L+M_{e}\Pi\left(\theta^{\ast},M_{e},\lambda\right)\right],
\end{equation*}
where $\Pi\left(\theta^{\ast},M_{e},\lambda\right)$ is the expected present value of entry net of entry costs:
\begin{equation*}
\Pi\left(\theta^{\ast},M_{e},\lambda\right)\coloneq\int_{\theta^{\ast}}^{\overline{\theta}}\frac{\widetilde{\pi}\left(\theta;\varphi^{\ast}\!\left(\lambda\right)\right)}{\delta}dC\left(\theta\right)-\left[1-C\left(\theta^{\ast}\right)\right]f_{b}\left(M_{e}\right)-f_{n}\left(M_{e}\right).
\end{equation*}
The productivity threshold is henceforth denoted with $\varphi^{\ast}\!\left(\lambda\right)$ to make its dependence on $\lambda$ at stage-three explicit. Through the chain and Leibniz rules, one can also show that:
\begin{equation*}
\frac{\partial\Pi\left(\theta^{\ast},M_{e},\lambda\right)}{\partial\lambda}=-\frac{\sigma f}{\lambda\delta}\int_{\theta^{\ast}}^{\overline{\theta}}\int_{\varphi^{\ast}\!\left(\lambda\right)}^{\infty}\left(\frac{\varphi}{\varphi^{\ast}\!\left(\lambda\right)}\right)^{\sigma-1}\,dG\left(\varphi\mid\theta\right)\,dC\left(\theta\right),
\end{equation*}
using $\pi\left(\varphi^{\ast};\varphi^{\ast}\right)=0$ as well as:
\begin{equation*}
\frac{\partial\varphi^{\ast}\!\left(\lambda\right)}{\partial\lambda}=\frac{\sigma}{\sigma-1}\frac{\varphi^{\ast}\!\left(\lambda\right)}{\lambda}.
\end{equation*}

The key statements of this proposition are proven via the First Order Conditions (FOCs) of the reduced Lagrangian with respect to $\theta^{\ast}$ and $M_{e}$, which are as follows.
\begin{align*}
\frac{\partial\mathcal{L}\left(\theta^{\ast},M_{e},\lambda\right)}{\partial\theta^{\ast}}&=-\lambda M_{e}\left[\frac{\widetilde{\pi}\left(\theta^{\ast};\varphi^{\ast}\!\left(\lambda\right)\right)}{\delta}-f_{b}\left(M_{e}\right)\right]\frac{dC\left(\theta^{\ast}\right)}{d\theta}=0\\
\frac{\partial\mathcal{L}\left(\theta^{\ast},M_{e},\lambda\right)}{\partial M_{e}}&=\lambda\left[\int_{\theta^{\ast}}^{\overline{\theta}}\frac{\widetilde{\pi}\left(\theta;\varphi^{\ast}\!\left(\lambda\right)\right)}{\delta}dC\left(\theta\right)-\left[1-C\left(\theta^{\ast}\right)\right]\left[1+\varepsilon_{b}\left(M_{e}\right)\right]f_{b}\left(M_{e}\right)-\left[1+\varepsilon_{n}\left(M_{e}\right)\right]f_{n}\left(M_{e}\right)\right]=0
\end{align*}
Their analysis reveals that any interior equilibrium with overall entry $M_{e}$ is socially optimal, provided that $\tau_{n}^{\circ}=0$. In fact, in this case the two FOCs reduce to the AC and FE conditions, respectively. Conversely, if an equilibrium is a social optimum it must feature $\tau_{n}^{\circ}=0$ for the two FOCs to coincide with the decentralized AC and FE conditions. To implement the constrained-efficient allocation when $\tau_{n}^{\circ}\neq0$, a policymaker can set stage-one Pigouvian taxes or subsidies $\tau_{n}=\tau_{n}^{\circ}$, while leaving $\tau_{b}=0$ at stage two. However, no budget-neutral pair of instruments $\smash{\left(\tau_{n},\tau_{b}\right)}$ is available to the policymaker, since $\tau_{b}\neq0$ would inevitably introduce a wedge between the private AC (with taxes) and the social AC. To complete the analysis, it is useful to derive the FOC with respect to $\lambda$ via the chain rule and the derivative of $\Pi\left(\theta^{\ast},M_{e},\lambda\right)$ with respect to $\lambda$:
\begin{multline*}
\frac{\partial\mathcal{L}\left(\theta^{\ast},M_{e},\lambda\right)}{\partial\lambda}=L-M_{e}\left[\left(\sigma-1\right)\frac{f}{\delta}\int_{\theta^{\ast}}^{\overline{\theta}}\int_{\varphi^{\ast}\!\left(\lambda\right)}^{\infty}\left(\frac{\varphi}{\varphi^{\ast}\!\left(\lambda\right)}\right)^{\sigma-1}\,dG\left(\varphi\mid\theta\right)\,dC\left(\theta\right)\right.+\\+\left.\vphantom{\left(\frac{\varphi}{\varphi^{\ast}\!\left(\lambda\right)}\right)^{\sigma-1}}\frac{f}{\delta}p(\theta^{\ast},\varphi^{\ast}\!\left(\lambda\right))+\left[1-C\left(\theta^{\ast}\right)\right]f_{b}\left(M_{e}\right)+f_{n}\left(M_{e}\right)\right]=0.
\end{multline*}
This FOC shows that the labor constraint binds with equality at every solution.
\end{proof}

\begin{proof}[Proof of Proposition \ref{prop:neutrality}] This is a minor variation of the proof of Proposition \ref{prop:breakdown}. The planner maximizes the utility of the representative consumer under the labor resource constraints, anticipating that for each successful variety $\omega$, quantities $q$ are set optimally, and production occurs only if the observed productivity exceeds the resulting cutoff $\varphi^{\ast}$. The reduced Lagrangian is still:
\begin{equation*}
\mathcal{L}\left(\theta^{\ast},M_{e},\lambda\right)=\lambda\left[L+M_{e}\Pi\left(\theta^{\ast},M_{e},\lambda\right)\right],
\end{equation*}
but the function expressing the net expected present value of entry now reads:
\begin{equation*}
\Pi\left(\theta^{\ast},M_{e},\lambda\right)\coloneq\int_{\theta^{\ast}}^{\overline{\theta}}\frac{\widetilde{\pi}\left(\theta;\varphi^{\ast}\!\left(\lambda\right)\right)}{\delta}dC\left(\theta\right)-\left[1-C\left(\theta^{\ast}\right)\right]f_{b}\left(\left[1-C\left(\theta^{\ast}\right)\right]M_{e}\right)-f_{n}\left(M_{e}\right).
\end{equation*}
With respect to the proof of Proposition \ref{prop:breakdown}, $f_{b}\left(M_{n}\right)$ now replaces $f_{b}\left(M_{e}\right)$, with $M_{n}=M_{e}\left[1-C\left(\theta^{\ast}\right)\right]$. The relevant FOCs with respect to $\theta^{\ast}$ and $M_{e}$ are now as follows.
\begin{align*}
\frac{\partial\mathcal{L}\left(\theta^{\ast},M_{e},\lambda\right)}{\partial\theta^{\ast}}&=-\lambda M_{e}\left[\frac{\widetilde{\pi}\left(\theta^{\ast};\varphi^{\ast}\!\left(\lambda\right)\right)}{\delta}-\left[1+\varepsilon_{b}\left(M_{n}\right)\right]f_{b}\left(M_{n}\right)\right]\frac{dC\left(\theta^{\ast}\right)}{d\theta}=0\\
\frac{\partial\mathcal{L}\left(\theta^{\ast},M_{e},\lambda\right)}{\partial M_{e}}&=\lambda\left[\int_{\theta^{\ast}}^{\overline{\theta}}\frac{\widetilde{\pi}\left(\theta;\varphi^{\ast}\!\left(\lambda\right)\right)}{\delta}dC\left(\theta\right)-\left[1-C\left(\theta^{\ast}\right)\right]\left[1+\varepsilon_{b}\left(M_{n}\right)\right]f_{b}\left(M_{n}\right)-\left[1+\varepsilon_{n}\left(M_{e}\right)\right]f_{n}\left(M_{e}\right)\right]=0    
\end{align*}
The inspection of both FOCs reveals that if $\tau_{n}^{\ast}=\tau_{b}^{\ast}=0$, they reduce to the AC and FE conditions, respectively; and the Pigouvian instruments specified in the statement align the incentives of decentralized entrants with the planner's. The resulting socially optimal decentralized equilibrium with Pigouvian instruments complies with budget neutrality if $\tau_{n}^{\ast}+[1-C\left(\theta^{\ast}\right)]\tau_{b}^{\ast}=0$. Simplifying this expression leads to the budget neutrality condition stated in the proposition. As in Proposition \ref{prop:breakdown}, the FOC with respect to $\lambda$ shows that the labor constraint binds at every solution.
\end{proof}

\clearpage\section*{Calibration: additional details and results}

This appendix reports additional details and results related to the quantitative exercise of Section \ref{sec:calibration}. It is organized as three self-contained parts: the first reconstructs the productivity variance moment used in the calibration; the second develops the analytical formulae used in calculations; the third illustrates additional results about the optimal Pigouvian instruments. Throughout, we use $x\coloneq\log\theta$ and $y\coloneq\log\varphi$ to simplify notation. In logarithms, the two threshold pairs are thus expressed as $\log\theta^{\ast}=x^{\ast}$ and $\log\varphi^{\ast}=y^{\ast}$.

\subsection*{Reconstruction of the productivity variance}

We derive the fifth moment of our calibration from extant estimates of \emph{within-industry} dispersion of Total Factor Productivity (TFP) by \citet{syverson2004product,syverson2011what} and \citet{foster2008reallocation}. We consider (physical) TFP an appropriate proxy for our model's productivity variable $\varphi$. It is important to remark that this calculation ignores variation \emph{between} industries. Our model does not distinguish between industries, and the BFS data we use in our quantitative exercise span multiple sectors: hence, the fifth moment must be considered a conservative lower bound. Increasing the target variance further would shrink $\psi$ in our calibration, leading to fatter productivity tails and higher welfare gains from stronger incentives to experimentation.

Here we show that the estimates from the literature that we use are consistent with one another, and with our model. \citet{syverson2004product,syverson2011what} reports that the interdecile 90-10 ratio of within-industry TFP (in levels) is approximately 1.92, while in their Table 1, \citet{foster2008reallocation} report the standard deviation of \emph{logarithmic} within-industry TFP as 0.26. We reconcile both figures with the truncated distribution of $y$ implied by both our model and our empirical calibration. To proceed, recall that $y\sim\mathcal{N}\left(\mu,1/\psi^{2}\right)$, and hence, $z\coloneq\psi\left(y-\mu\right)\sim\mathcal{N}\left(0,1\right)$. Define firm survival in the baseline equilibrium where $y^{\ast}=0$ as the event:
\begin{equation*}
S=\left\{x\geq x^{\ast}\!,z\geq-\mu\psi\right\}.
\end{equation*}
Note that $x\mid z\sim\mathcal{N}\left(\rho z,1-\rho^{2}\right)$. From Bayes' rule, the density of $z$ among operating firms (i.e., conditional on $S$), which with some slight abuse of notation we incidentally denote by $f_{z\mid S}\left(z\right)$, is
\begin{equation*}
f_{z\mid S}\left(z\right)=\frac{\phi\left(z\right)\Prob\left(S\mid z\right)}{\Prob\left(S\right)}=\frac{\phi\left(z\right)\Prob\left(x\geq x^{\ast}\mid z\right)\mathds{1}\left\{z\geq-\mu\psi\right\}}{\Prob\left(x\geq x^{\ast}\!,z\geq-\mu\psi\right)}=\frac{\phi\left(z\right)\left[1-\Phi
\left(\frac{x^\ast-\rho z}{\sqrt{1-\rho^2}}
\right)\right]}{\overline{\Phi}\left(x^{\ast}\!,-\mu\psi;\rho\right)}
\mathds{1}\left\{z\geq-\mu\psi\right\},
\end{equation*}
where $\phi\left(\cdot\right)$, not to be confused with $\varphi$, is the density function of the standard normal distribution. Under the calibration values $x^{\ast}=0.018$, $\rho=0.644$ and $\mu\psi=-1.036$, numerical integration over (moments of) $\smash{f_{z\mid S}\left(z\right)}$ allows to calculate the conditional variance
\begin{equation*}
\mathbb{V}\mathrm{ar}\left[z\mid S\right]=0.203,
\end{equation*}
and to retrieve the quantiles of $z\mid S$. From the latter, one can easily calculate the interdecile range of $z\mid S$ (the difference between the 90th and 10th percentiles) as
\begin{equation*}
IDC\left(z\mid S\right)=1.099.    
\end{equation*}
Further observe that, because $y$ and $z$ are affine transformations of one another:
\begin{equation*}
IDC\left(y\mid S\right)=\frac{IDC\left(z\mid S\right)}{\psi}=\log\left(1.92\right)
\end{equation*}
where 1.92 is the actual value of the interdecile \emph{ratio} for the \emph{levels} of TFP as in \citet{syverson2004product,syverson2011what}. Thus:
\begin{equation*}
\mathbb{V}\mathrm{ar}\left[y\mid S\right]=\frac{\mathbb{V}\mathrm{ar}\left[z\mid S\right]}{\psi^{2}}=\mathbb{V}\mathrm{ar}\left[z\mid S\right]\cdot\left(\frac{\log\left(1.92\right)}{IDC\left(z\mid S\right)}\right)^{2}=0.203\cdot\left(\frac{0.652}{1.099}\right)^{2}=0.071,
\end{equation*}
implying a standard deviation of $y\mid S$ around 0.266. This is extremely close to the direct estimate by \citet{foster2008reallocation}, which implies
\begin{equation*}
\mathbb{V}\mathrm{ar}\left[y\mid S\right]=0.067.    
\end{equation*}
The fifth moment we use in the calibration is a central value between these two conditional variance figures.

\subsection*{Profit and equilibrium equations}

We next develop the key equations of our quantitative exercise under our bivariate (log-)normality assumption. In particular, we characterize profits as a function of a signal $x$, productivity $y$ and their thresholds, and the conditions characterizing both the decentralized and social equilibria. Under CES demand, and holding the normalization $f=1$, the flow profit of a firm that has learned its normally-distributed log-productivity $y$ is:
\begin{equation*}
\pi\left(y;y^{\ast}\!\right)=\left[\exp\left(\left(\sigma-1\right)\left(y-y^{\ast}\!\right)\right)-1\right]\mathds{1}\left\{y\geq y^{\ast}\!\right\}.
\end{equation*}
The conditional distribution of log-productivity $y$ given a log-signal $x$ is normal: $y\mid x\sim\mathcal{N}\left(\nu\left(x\right),\xi^{2}\right)$, where:
\begin{equation*}
\nu\left(x\right)\coloneq\mu+\frac{\rho}{\psi}x\quad\text{and}\quad\xi\coloneq\frac{\sqrt{1-\rho^{2}}}{\psi}.
\end{equation*}
Using established results about the truncated lognormal distribution, one can thus calculate the flow profit that an entrant expects after observing $x$ but before observing $y$, i.e., the version of \eqref{eq:pi-tilde} used in the exercise:
\begin{equation*}
\widetilde{\pi}\left(x;y^{\ast}\!\right)=\mathbb{E}\left[\pi\left(y;y^{\ast}\!\right)\mid x\right]=
\exp\left(\widetilde{\sigma}\left(\nu\left(x\right)-y^{\ast}\!\right)+\frac{\widetilde{\sigma}^{2}\xi^{2}}{2}\right)\Phi\left(\frac{\nu\left(x\right)-y^{\ast}\!+\widetilde{\sigma}\xi^{2}}{\xi}\right)-\Phi\left(\frac{\nu\left(x\right)-y^{\ast}}{\xi}\right),
\end{equation*}
where $\widetilde{\sigma}\coloneq\sigma-1$. Under the normalization $y^{\ast}=0$ applied to the calibrated baseline, the AC condition reads:
\begin{equation*}
\frac{\widetilde{\pi}\left(0.018;0\right)}{\delta}-f_{b}^{0}=0,
\end{equation*}
which allows to retrieve the baseline stage-two entry cost. Next, let the expected flow of \emph{variable} profits be
\begin{equation*}
h\left(x^{\ast}\!,y^{\ast}\!\right)\coloneq
\mathbb{E}\left[\exp\left(\widetilde{\sigma}\left(y-y^{\ast}\!\right)\right)\mathds{1}\left\{x\geq x^{\ast}\!,y\geq y^{\ast}\!\right\}\right]=\exp\left(\widetilde{\sigma}\left(\mu-y^{\ast}\!\right)+\frac{\widetilde{\sigma}^{2}}{2\psi^{2}}\right)\overline{\Phi}\left(x^{\ast}-\frac{\rho\widetilde{\sigma}}{\psi},\psi\left(y^{\ast}-\mu\right)-\frac{\widetilde{\sigma}}{\psi};\rho\right),
\end{equation*}
where the expectation is taken over the unconditional joint distribution of $\left(x,y\right)$, and is developed using standard properties of the bivariate normal distribution. Furthermore, by slightly abusing some notation already introduced in the proofs of Proposition \ref{prop:breakdown} and \ref{prop:neutrality}, we express the expected (out)flow of fixed costs as:
\begin{equation*}
p\left(x^{\ast}\!,y^{\ast}\!\right)\coloneq
\Prob\left(x\geq x^{\ast}\!,y\geq y^{\ast}\!\right)=\overline{\Phi}\left(x^{\ast}\!,\psi\left(y^{\ast}-\mu\right);\rho\right).
\end{equation*}
Holding the normalization $y^{\ast}=0$, the FE condition in the calibrated equilibrium is thus:
\begin{equation*}
\frac{h\left(0.018;0\right)-p\left(0.018;0\right)}{\delta}-\left[1-\Phi\left(0.018\right)\right]f_{b}^{0}-f_{n}^{0}=0,
\end{equation*}
which allows to solve for the baseline stage-one entry costs. In addition, one can retrieve total labor $L$ as:
\begin{equation*}
L=\frac{\left(\sigma-1\right)h\left(0.018;0\right)+p\left(0.018;0\right)}{\delta}+\left[1-\Phi\left(0.018\right)\right]f_{b}^{0}+f_{n}^{0}.
\end{equation*}
We construct Figure \ref{fig:calibration_exercise} by solving for triples $\left(x^{\ast}\!,y^{\ast}\!,M_{e}\right)$ that satisfy the social planner's optimality conditions developed in Proposition \ref{prop:neutrality}: the ``social'' AC and FE conditions, and the binding resource constraint:
\begin{align*}
\frac{\widetilde{\pi}\left(x^{\ast};y^{\ast}\!\right)}{\delta}-\left[1+\varepsilon_{b}\left(M_{n}\right)\right]f_{b}\left(M_{n}\right)&=0\\
\frac{h\left(x^{\ast};y^{\ast}\!\right)-p\left(x^{\ast};y^{\ast}\!\right)}{\delta}-\left[1-\Phi\left(x^{\ast}\!\right)\right]\left[1+\varepsilon_{b}\left(M_{n}\right)\right]f_{b}\left(M_{n}\right)-\left[1+\varepsilon_{n}\left(M_{e}\right)\right]f_{n}\left(M_{e}\right)&=0\\
L-M_{e}\left[\frac{\left(\sigma-1\right)h\left(x^{\ast};y^{\ast}\!\right)+p\left(x^{\ast};y^{\ast}\!\right)}{\delta}+\left[1-\Phi\left(x^{\ast}\!\right)\right]f_{b}\left(M_{n}\right)+f_{n}\left(M_{e}\right)\right]&=0
\end{align*}
where $M_{n}=\left[1-\Phi\left(x^{\ast}\!\right)\right]M_{e}$, while $y^{\ast}$ generally differs from zero. The equilibrium firm mass and aggregate productivity displayed in the two panels of the figure are then calculated as:
\begin{equation*}
M=\frac{M_{e}}{\delta}p\left(x^{\ast};y^{\ast}\!\right)\quad\text{and}\quad\widetilde{\varphi}\left(x^{\ast};y^{\ast}\!\right)=\exp\left(y^{\ast}\!\right)\left[\frac{h\left(x^{\ast};y^{\ast}\!\right)}{p\left(x^{\ast};y^{\ast}\!\right)}\right]^{\displaystyle\nicefrac{1}{\sigma-1}}.
\end{equation*}

\subsection*{Additional calibration results}

We conclude this appendix with a report and accompanying discussion of additional results from our quantitative exercise. We sequentially discuss: \begin{enumerate*}[label=\roman*., itemjoin={{; }}, itemjoin*={{; and }}] \item the alternative calibration using BF4Q \item results based on $\sigma\in\left\{3,7\right\}$ \item aggregate paid entry costs across the social equilibria of Figure \ref{fig:calibration_exercise}\end{enumerate*}. All figures report percentage changes relative to the implied calibrated baseline and display separate curves for each of the three reference values of the spillover elasticity. All horizontal axes run through the congestion elasticity interval $\varepsilon_{b}^{0}\in\left[0.0,0.1\right]$.

\paragraph{Alternative calibration} The alternative calibration mentioned in the main text redefines successful entry as BF4Q (business formation after four quarters) status from the BFS data, instead of BF8Q (eight quarters). This changes the two key moments that express the probability of successful entry conditional on attempting the first stage; these are recast as follows.
\begin{align*}
\Prob\left(\text{BF4Q}\cap\text{HBA}\mid\text{BA}\right)&=\overline{\Phi}\left(\log\theta^{\ast},-\mu\psi;\rho\right)=0.117\\
\Prob\left(\text{BF4Q}\cap\text{WBA}\mid\text{BA}\right)&=\overline{\Phi}\left(\log\theta^{w},-\mu\psi;\rho\right)=0.087
\end{align*}
This implies only minor changes to the calibrated parameters: now, $\rho=0.670$, $\mu=-0.684$, and $\psi=1.661$. Maintaining $\sigma=5$ and $\delta=0.02$, the two entry costs at the baseline equilibrium are $f_{n}^{0}=42.296$ and $f_{b}^{0}=6.775$, while total labor is solved as $L=257.430$. Figure \ref{fig:appendix-calibration-bf4q} reports the changes in $M$ and $\widetilde{\varphi}\left(x^{\ast},y^{\ast}\right)$ for all elasticity pairs $\smash{\left(\varepsilon_{n}^{0},\varepsilon_{b}^{0}\right)}$ examined in Figure \ref{fig:calibration_exercise}. Unsurprisingly, the qualitative patterns are identical, as quantitatively the alternative calibration only yields minor differences.

\begin{figure}[ht!]
    \centering
    \caption{Calibrated optimal policy using BF4Q}
    \label{fig:appendix-calibration-bf4q}
    \begin{tikzpicture}
    \begin{groupplot}[
        group style={group size=2 by 1, horizontal sep=1cm},
        axis lines=left,
        axis line style={->},
        scale only axis,
        width=0.405\textwidth,
        height=4.85cm,
        xmin=0,
        xmax=0.100,
        enlarge x limits={upper},
        enlarge y limits={upper},
        xtick={0,0.02,0.04,0.06,0.08,0.10},
        xticklabels={0.00,0.02,0.04,0.06,0.08,0.10},
        scaled x ticks=false,
        xlabel={Congestion elasticity $\varepsilon_{b}^{0}$},
        ylabel={Percentage change},
        label style={font=\footnotesize},
        tick label style={font=\footnotesize},
        legend style={font=\footnotesize, fill=white, draw=black!35, rounded corners=0pt, inner sep=2pt, /tikz/every even column/.append style={column sep=0.35cm}},
        clip=true]
        \nextgroupplot[
            title={A. Operating firm mass},
            title style={font=\small},
            ymin=-0.4,
            ymax=1.3,
            ytick={-0.4,0,0.4,0.8,1.2},
            legend to name=appendixbf4qlegend,
            legend columns=3]
            \addplot[very thick, black!55, densely dotted]
                table[x=epsilon_b,y=firmmass_eps003, col sep=comma] {figures/appendix_bf4q_panel_a.csv};
            \addlegendentry{$\varepsilon_n^0=-0.03$}
            \addplot[very thick, black]
                table[x=epsilon_b,y=firmmass_eps005, col sep=comma] {figures/appendix_bf4q_panel_a.csv};
            \addlegendentry{$\varepsilon_n^0=-0.05$}
            \addplot[very thick, black!70, dashed]
                table[x=epsilon_b,y=firmmass_eps008, col sep=comma] {figures/appendix_bf4q_panel_a.csv};
            \addlegendentry{$\varepsilon_n^0=-0.08$}
        \nextgroupplot[
            title={B. Aggregate productivity},
            title style={font=\small},
            ymin=0,
            ymax=1.4,
            ytick={0,0.4,0.8,1.2},
            ylabel={}]
            \addplot[very thick, black!55, densely dotted]
                table[x=epsilon_b,y=productivity_eps003, col sep=comma] {figures/appendix_bf4q_panel_b.csv};
            \addplot[very thick, black]
                table[x=epsilon_b,y=productivity_eps005, col sep=comma] {figures/appendix_bf4q_panel_b.csv};
            \addplot[very thick, black!70, dashed]
                table[x=epsilon_b,y=productivity_eps008, col sep=comma] {figures/appendix_bf4q_panel_b.csv};
    \end{groupplot}
    \node[anchor=north, yshift=-1.35cm] at ($(group c1r1.south east)!0.5!(group c2r1.south west)$) {\pgfplotslegendfromname{appendixbf4qlegend}};
    \end{tikzpicture}
    \note[Note]{The figure repeats Figure \ref{fig:calibration_exercise} after recalibrating successful entry using business formations within four quarters rather than within eight quarters. The key BF4Q moments are $\Prob\left(\text{BF4Q}\cap\text{HBA}\mid\text{BA}\right)=0.117$ and $\Prob\left(\text{BF4Q}\cap\text{WBA}\mid\text{BA}\right)=0.087$.}
\end{figure}

\paragraph{Perturbed substitution elasticities} Our calibration exercise maintains $\sigma=5$, a value associated with a moderate and realistic markup (which is constant across firms). We thus repeat the exercise under two alternative values of $\sigma$, one smaller and one larger:
\begin{itemize}[itemsep=6pt,topsep=6pt]
    \item $\sigma=3$, for a markup of 1.5, yielding baseline entry costs $f_{n}^{0}=6.855$ and $f_{b}^{0}=2.861$, and total labor $L=44.763$;
    \item $\sigma=7$, for a markup of 1.166, yielding baseline entry costs $f_{n}^{0}=559.166$ and $f_{b}^{0}=40.571$, and total labor $L=4100.760$.
\end{itemize}

\begin{figure}[ht!]
    \centering
    \caption{Calibrated optimal policy for $\sigma=3$}
    \label{fig:appendix-calibration-sigma3}
    \begin{tikzpicture}
    \begin{groupplot}[
        group style={group size=2 by 1, horizontal sep=1cm},
        axis lines=left,
        axis line style={->},
        scale only axis,
        width=0.405\textwidth,
        height=4.85cm,
        xmin=0,
        xmax=0.100,
        enlarge x limits={upper},
        enlarge y limits={upper},
        xtick={0,0.02,0.04,0.06,0.08,0.10},
        xticklabels={0.00,0.02,0.04,0.06,0.08,0.10},
        scaled x ticks=false,
        xlabel={Congestion elasticity $\varepsilon_{b}^{0}$},
        ylabel={Percentage change},
        label style={font=\footnotesize},
        tick label style={font=\footnotesize},
        legend style={font=\footnotesize, fill=white, draw=black!35, rounded corners=0pt, inner sep=2pt, /tikz/every even column/.append style={column sep=0.35cm}},
        clip=true]
        \nextgroupplot[
            title={A. Operating firm mass},
            title style={font=\small},
            ymin=-0.8,
            ymax=0.05,
            ytick={-0.6,-0.3,0},
            legend to name=appendixsigma3legend,
            legend columns=3]
            \addplot[very thick, black!55, densely dotted]
                table[x=epsilon_b,y=firmmass_eps003, col sep=comma] {figures/appendix_sigma3_panel_a.csv};
            \addlegendentry{$\varepsilon_n^0=-0.03$}
            \addplot[very thick, black]
                table[x=epsilon_b,y=firmmass_eps005, col sep=comma] {figures/appendix_sigma3_panel_a.csv};
            \addlegendentry{$\varepsilon_n^0=-0.05$}
            \addplot[very thick, black!70, dashed]
                table[x=epsilon_b,y=firmmass_eps008, col sep=comma] {figures/appendix_sigma3_panel_a.csv};
            \addlegendentry{$\varepsilon_n^0=-0.08$}
        \nextgroupplot[
            title={B. Aggregate productivity},
            title style={font=\small},
            ymin=0,
            ymax=1.8,
            ytick={0,0.6,1.2,1.8},
            ylabel={}]
            \addplot[very thick, black!55, densely dotted]
                table[x=epsilon_b,y=productivity_eps003, col sep=comma] {figures/appendix_sigma3_panel_b.csv};
            \addplot[very thick, black]
                table[x=epsilon_b,y=productivity_eps005, col sep=comma] {figures/appendix_sigma3_panel_b.csv};
            \addplot[very thick, black!70, dashed]
                table[x=epsilon_b,y=productivity_eps008, col sep=comma] {figures/appendix_sigma3_panel_b.csv};
    \end{groupplot}
    \node[anchor=north, yshift=-1.35cm] at ($(group c1r1.south east)!0.5!(group c2r1.south west)$) {\pgfplotslegendfromname{appendixsigma3legend}};
    \end{tikzpicture}
    \note[Note]{The figure is a version of Figure \ref{fig:calibration_exercise} obtained under the assumption $\sigma=3$; recalibrating both entry costs at the decentralized baseline while holding the same entry and productivity moments.}
\end{figure}

\begin{figure}[ht!]
    \centering
    \caption{Calibrated optimal policy for $\sigma=7$}
    \label{fig:appendix-calibration-sigma7}
    \begin{tikzpicture}
    \begin{groupplot}[
        group style={group size=2 by 1, horizontal sep=1cm},
        axis lines=left,
        axis line style={->},
        scale only axis,
        width=0.405\textwidth,
        height=4.85cm,
        xmin=0,
        xmax=0.100,
        enlarge x limits={upper},
        enlarge y limits={upper},
        xtick={0,0.02,0.04,0.06,0.08,0.10},
        xticklabels={0.00,0.02,0.04,0.06,0.08,0.10},
        scaled x ticks=false,
        xlabel={Congestion elasticity $\varepsilon_{b}^{0}$},
        ylabel={Percentage change},
        label style={font=\footnotesize},
        tick label style={font=\footnotesize},
        legend style={font=\footnotesize, fill=white, draw=black!35, rounded corners=0pt, inner sep=2pt, /tikz/every even column/.append style={column sep=0.35cm}},
        clip=true]
        \nextgroupplot[
            title={A. Operating firm mass},
            title style={font=\small},
            ymin=-0.2,
            ymax=3.4,
            ytick={0,1,2,3},
            legend to name=appendixsigma7legend,
            legend columns=3]
            \addplot[very thick, black!55, densely dotted]
                table[x=epsilon_b,y=firmmass_eps003, col sep=comma] {figures/appendix_sigma7_panel_a.csv};
            \addlegendentry{$\varepsilon_n^0=-0.03$}
            \addplot[very thick, black]
                table[x=epsilon_b,y=firmmass_eps005, col sep=comma] {figures/appendix_sigma7_panel_a.csv};
            \addlegendentry{$\varepsilon_n^0=-0.05$}
            \addplot[very thick, black!70, dashed]
                table[x=epsilon_b,y=firmmass_eps008, col sep=comma] {figures/appendix_sigma7_panel_a.csv};
            \addlegendentry{$\varepsilon_n^0=-0.08$}
        \nextgroupplot[
            title={B. Aggregate productivity},
            title style={font=\small},
            ymin=0,
            ymax=1.0,
            ytick={0,0.4,0.8},
            ylabel={}]
            \addplot[very thick, black!55, densely dotted]
                table[x=epsilon_b,y=productivity_eps003, col sep=comma] {figures/appendix_sigma7_panel_b.csv};
            \addplot[very thick, black]
                table[x=epsilon_b,y=productivity_eps005, col sep=comma] {figures/appendix_sigma7_panel_b.csv};
            \addplot[very thick, black!70, dashed]
                table[x=epsilon_b,y=productivity_eps008, col sep=comma] {figures/appendix_sigma7_panel_b.csv};
    \end{groupplot}
    \node[anchor=north, yshift=-1.35cm] at ($(group c1r1.south east)!0.5!(group c2r1.south west)$) {\pgfplotslegendfromname{appendixsigma7legend}};
    \end{tikzpicture}
    \note[Note]{The figure is a version of Figure \ref{fig:calibration_exercise} obtained under the assumption $\sigma=7$; recalibrating both entry costs at the decentralized baseline while holding the same entry and productivity moments.}
\end{figure}

\noindent The implications on the two determinants of welfare in the social optima are illustrated in Figures \ref{fig:appendix-calibration-sigma3} and \ref{fig:appendix-calibration-sigma7}, respectively. Qualitatively, the results are very similar; quantitatively, however, a number of differences are noticeable. Higher markups ($\sigma=3$) magnify selection effects in a way that trades off variety $M$ in exchange for higher aggregate productivity $\widetilde{\varphi}\left(x^{\ast},y^{\ast}\right)$; however, the latter now \emph{declines} as congestion effects becomes stronger. Lower markups ($\sigma=7$), by contrast, deliver meaningful variety increases for higher spillover forces and, at the same time, weaker productivity effects.

\paragraph{Changes in paid entry costs} Lastly, we examine how total paid entry costs change in the social equilibria represented in Figure 4. Let $\smash{\tau_{n}\left(\varepsilon_{n}^{0},\varepsilon_{b}^{0}\right)}$ and $\smash{\tau_{b}\left(\varepsilon_{n}^{0},\varepsilon_{b}^{0}\right)}$ be, respectively, the stage-one and stage-two Pigouvian instruments associated with the social optimum identified by a given elasticity pair $\smash{\left(\varepsilon_{n}^{0},\varepsilon_{b}^{0}\right)}$. Analogously, let $M_{e}\left(\varepsilon_{n}^{0},\varepsilon_{b}^{0}\right)$ and $M_{n}\left(\varepsilon_{n}^{0},\varepsilon_{b}^{0}\right)$ be the stage-one and stage-two active entrants in the same social optima. For each social equilibrium, we calculate the \emph{overall} changes in \emph{total} entry costs (relative to the decentralized baseline) that are paid in either stage as the ratios
\begin{equation*}
\frac{M_{e}\left(\varepsilon_{n}^{0},\varepsilon_{b}^{0}\right)\left[f_{n}\left(M_{e}\left(\varepsilon_{n}^{0},\varepsilon_{b}^{0}\right)\right)+\tau_{n}\left(\varepsilon_{n}^{0},\varepsilon_{b}^{0}\right)\right]}{M_{e}^{0}f_{n}^{0}}-1\quad\text{and}\quad
\frac{M_{n}\left(\varepsilon_{n}^{0},\varepsilon_{b}^{0}\right)\left[f_{b}\left(M_{n}\left(\varepsilon_{n}^{0},\varepsilon_{b}^{0}\right)\right)+\tau_{b}\left(\varepsilon_{n}^{0},\varepsilon_{b}^{0}\right)\right]}{M_{n}^{0}f_{b}^{0}}-1,
\end{equation*}
and we recast them as percentages. The results are displayed in Figure \ref{fig:appendix-calibration-costs}: Panels A and B report, respectively, the overall variations of first-stage and second-stage entry costs. The former decrease only mildly (between 1 and 2 percent), as subsidies are partly offset by an increase in equilibrium entrants $M_{e}$. By contrast, the latter increase more markedly (between 2 and 8 percent), as taxes add up to a rise in $M_{n}$, only partly mitigated by sharper selection (higher $x^{\ast}$). Some back-of-the-envelope calculations show that the net fiscal cost of the Pigouvian policy is positive (total subsidies surpass tax revenues) across all social equilibria we identify, ranging from about 0.31 to 1.33 percent of the total labor mass $L$.

\begin{figure}[ht!]
    \centering
    \caption{Calibrated optimal policy: changes in total entry costs}
    \label{fig:appendix-calibration-costs}
    \begin{tikzpicture}
    \begin{groupplot}[
        group style={group size=2 by 1, horizontal sep=1cm},
        axis lines=left,
        axis line style={->},
        scale only axis,
        width=0.405\textwidth,
        height=4.85cm,
        xmin=0,
        xmax=0.100,
        enlarge x limits={upper},
        enlarge y limits={upper},
        xtick={0,0.02,0.04,0.06,0.08,0.10},
        xticklabels={0.00,0.02,0.04,0.06,0.08,0.10},
        scaled x ticks=false,
        xlabel={Congestion elasticity $\varepsilon_{b}^{0}$},
        ylabel={Percentage change},
        label style={font=\footnotesize},
        tick label style={font=\footnotesize},
        legend style={font=\footnotesize, fill=white, draw=black!35, rounded corners=0pt, inner sep=2pt, /tikz/every even column/.append style={column sep=0.35cm}},
        clip=true]
        \nextgroupplot[
            title={A. Total first-stage payments $M_{e}f_{n}$},
            title style={font=\small},
            ymin=-3,
            ymax=0,
            ytick={-3,-2,-1,0},
            legend to name=appendixcostslegend,
            legend columns=3]
            \addplot[very thick, black!55, densely dotted]
                table[x=epsilon_b,y=firsttotal_eps003, col sep=comma] {figures/appendix_costs_panel_a.csv};
            \addlegendentry{$\varepsilon_n^0=-0.03$}
            \addplot[very thick, black]
                table[x=epsilon_b,y=firsttotal_eps005, col sep=comma] {figures/appendix_costs_panel_a.csv};
            \addlegendentry{$\varepsilon_n^0=-0.05$}
            \addplot[very thick, black!70, dashed]
                table[x=epsilon_b,y=firsttotal_eps008, col sep=comma] {figures/appendix_costs_panel_a.csv};
            \addlegendentry{$\varepsilon_n^0=-0.08$}
        \nextgroupplot[
            title={B. Total second-stage payments $M_{n}f_{b}$},
            title style={font=\small},
            ymin=0,
            ymax=10,
            ytick={0,2,4,6,8,10},
            ylabel={}]
            \addplot[very thick, black!55, densely dotted]
                table[x=epsilon_b,y=secondtotal_eps003, col sep=comma] {figures/appendix_costs_panel_b.csv};
            \addplot[very thick, black]
                table[x=epsilon_b,y=secondtotal_eps005, col sep=comma] {figures/appendix_costs_panel_b.csv};
            \addplot[very thick, black!70, dashed]
                table[x=epsilon_b,y=secondtotal_eps008, col sep=comma] {figures/appendix_costs_panel_b.csv};
    \end{groupplot}
    \node[anchor=north, yshift=-1.35cm] at ($(group c1r1.south east)!0.5!(group c2r1.south west)$) {\pgfplotslegendfromname{appendixcostslegend}};
    \end{tikzpicture}
    \note[Note]{For all socially optimal equilibria of Figure \ref{fig:calibration_exercise}, Panel A reports the percentage change in total stage-one paid entry costs, while Panel B reports the percentage change in total stage-two paid entry costs (as per the formulae and discussion in this appendix).}
\end{figure}

\paragraph{Assessment} Taken together, the additional results point to a consistent and robust set of conclusions from our quantitative exercise. It is important to remark, however, that the results are based on a local search for social optima, as well as assumptions explicitly designed to make the exercise more transparent and tractable.

\end{document}